\documentclass[aps,prx,reprint,groupedaddress,floatfix]{revtex4-2}
\pdfoutput=1

\usepackage[T1]{fontenc}
\usepackage{newtxtext}
\usepackage{newtxmath}

\usepackage{amsmath}
\usepackage{mathtools}
\usepackage{bm}
\usepackage{bbold}
\usepackage{braket}

\DeclareMathOperator*{\argmax}{argmax}

\usepackage{color}
\usepackage{xcolor}
\usepackage[colorlinks,citecolor=blue,linkcolor=blue,urlcolor=blue]{hyperref}

\usepackage{graphicx}
\usepackage[all]{hypcap} 
\usepackage{multirow}

\begin{document}

\title{Spectral function of the $J_1-J_2$ Heisenberg model on the triangular lattice}

\author{Nicholas E. Sherman}
\affiliation{Department of Physics, University of California, Berkeley, California 94720, USA}
\affiliation{Materials Sciences Division, Lawrence Berkeley National Laboratory, Berkeley, California 94720, USA}

\author{Maxime Dupont}
\affiliation{Department of Physics, University of California, Berkeley, California 94720, USA}
\affiliation{Materials Sciences Division, Lawrence Berkeley National Laboratory, Berkeley, California 94720, USA}

\author{Joel E. Moore}
\affiliation{Department of Physics, University of California, Berkeley, California 94720, USA}
\affiliation{Materials Sciences Division, Lawrence Berkeley National Laboratory, Berkeley, California 94720, USA}

\begin{abstract}
    Spectral probes, such as neutron scattering, are crucial for characterizing excitations in quantum many-body systems and the properties of quantum materials. Among the most elusive phases of matter are quantum spin liquids, which have no long-range order even at zero temperature and host exotic fractionalized excitations with non-trivial statistics. These phases can occur in frustrated quantum magnets, such as the paradigmatic Heisenberg model with nearest and next-nearest neighbor exchange interactions on the triangular lattice, the so-called $J_1-J_2$ model. In this work, we compute the spectral function using large scale matrix product state simulations across the three different phases of this model's phase diagram, including a quantum spin liquid phase at intermediate $J_2/J_1$. Despite a plethora of theoretical and experimental studies, the exact nature of this phase is still contested, with the dominant candidates being a gapped $\mathbb{Z}_2$, a gapless $U(1)$ Dirac, and a spinon Fermi surface quantum spin liquid state. We find a V-shaped spectrum near the center of the Brillouin zone ($\Gamma$ point), a key signature of a spinon Fermi surface, observed in prior neutron scattering experiments. However, we find a small gap near the $\Gamma$ point, ruling out such a phase. Furthermore, we find localized gapless excitations at the corner of the Brillouin zone boundary (K point) and the middle of the edge of the Brillouin zone boundary (M point), ruling out the gapped $\mathbb{Z}_2$ spin liquid phase. Our results imply that the intermediate spin liquid phase is a gapless $U(1)$ Dirac spin liquid, and provide clear signatures to detect this phase in future neutron scattering experiments.
\end{abstract}

\maketitle

\section{Introduction}

Two-dimensional quantum systems host exciting physics: reduced dimensionality leads to strong quantum fluctuations, yet provides more possibilities than in one dimension as continuous symmetry can be spontaneously broken and lead to long-range order~\cite{PhysRevLett.17.1133,PhysRev.158.383,Momoi1996,giamarchi2004,sachdev2008}. While such conventional ordered states of matter are fairly well understood, some disordered states remain elusive. Among those are quantum spin liquids (QSL) found in frustrated quantum magnets~\cite{Savary2016,Knolle2019,Shaginyan2020,Broholm2020} as a result of competing ordered phases. These states possess no long-range order, even at zero temperature, and often result in fractionalized excitations with non-trivial statistics.

One of the most promising geometries for realizing a QSL phase is the triangular lattice, which has a rich history starting with Anderson's proposed resonating valence bond state~\cite{Anderson1973}. However, the simplest lattice spin model, namely the spin-$1/2$ nearest-neighbor antiferromagnetic Heisenberg model, has been shown to have 120$^{\circ}$ magnetic long-range ordering~\cite{Huse1988,Jolicoeur1989,Singh1992,Chubukov1994,Bernu1994,Capriotti1999,Zheng2006,White2007}. Yet, quantum fluctuations lead to the order parameter magnitude being significantly smaller than its classical value, implying the order is weak and potentially easy to disrupt. For instance, with the introduction of a small next-nearest-neighbor interaction, this model exhibits a QSL phase~\cite{PhysRevB.92.041105,PhysRevB.92.140403,Iqbal2016,PhysRevB.94.121111,PhysRevB.95.035141,PhysRevB.96.075116,Hu_2019}. Early studies using the density matrix renormalization group (DMRG)~\cite{White1992} suggested that the QSL phase was a $\mathbb{Z}_2$ gapped QSL~\cite{PhysRevB.92.041105,PhysRevB.92.140403, PhysRevB.94.121111}. This was later challenged by simulations using variational quantum Monte Carlo (QMC), which found that a gapless $U(1)$ Dirac spin liquid was most energetically favorable~\cite{Iqbal2016}. This was later supported by a DMRG study on an infinite cylinder with an external Aharonov-Bohm flux, claiming unambiguous evidence for a gapless $U(1)$ Dirac spin liquid~\cite{Hu_2019}. However, this has been challenged by a recent DMRG study \cite{Jiang_2022_arxiv}, as well as by Schwinger-boson theory~\cite{Ghioldi_2022_SBT}, suggesting the phase is a gapped $\mathbb{Z}_2$ QSL. 

The simplicity and realistic form of the Hamiltonian has attracted many experiments to probe triangular lattice materials, in the quest for a realization of such a QSL phase. Experiments conducted on triangular lattice systems range from organic compounds such as 
$k-$(BEDT-TTF)$_2$Cu$_2$(CN)$_3$, Et$_n$Me$_{4-n}$Sb[Pd(DMIT)$_2$]$_2$, and other similar structures
\cite{Yamashita2008, PhysRevB.89.045138, PhysRevLett.104.016403, PhysRevB.82.125119, PhysRevB.86.245103, PhysRevLett.117.107203, PhysRevB.90.195139, PhysRevB.86.155150, Nakajima2012, Furukawa2018, PhysRevB.85.134444, Yakushi_2015_JPS, Miksch_2021, PADMALEKHA2015211, PhysRevB.103.125111, PhysRevB.88.125101, PhysRevB.97.075115, cryst8020087, PhysRevResearch.2.042023, Muraoka_2011, Hartmann2019, PhysRevB.102.184417, PhysRevLett.104.016403,Itou2008, Itou_2009, PhysRevB.89.075105, PhysRevX.9.041051, PhysRevB.101.140407, cryst12010102, PhysRevB.105.245133,Yamashita_2019_JPS, Yamashita2022, Yamashita_2010_Science, PhysRevB.89.045113}, to Ba$_3$CoSb$_2$O$_9$ \cite{Shirata2012, Susuki2013, Naruse_2014, Koutroulakis2015, Quirion2015, Ma2016, Ito2017, Kamiya2018, Li2019, Macdougal2020, Zhang2020}, and many Yb$^{3+}$-based materials \cite{Li2015, Li_2015_YMGO, Li_2016_YMGO, Xu_2016_YMGO, Li2017, Paddison2017, Li2017_2, Li2017_3, Zhang2018, Baenitz_2018, Shen2018, Zhang_2021_NYS, Ding_2019_NYO, Bordelon2019, Ranjinth2019, Ranjith2019_2, sarkar2019quantum, Zangeneh2019, Li2019_YMGO, Sichelschmidt_2019, Sichelschmidt_2019, Xing_2019_Field-induced, Ding2020, Majumder2020, Xing2020, Bachus2020, Zeng2020, Guo2020, Zhang2020_arxiv, Pan2021, xing2021_KYS, Dai_2021, Xing2021, Ma2021, xie2021field, Rao2021, PhysRevLett.120.087201, Scheie2021Triangle}. In particular, recent neutron scattering data in KYbSe$_2$ has shown that the material is well modelled by a spin one-half Heisenberg model on a triangular lattice with nearest- and next-nearest-neighbor antiferromagnetic interactions, i.e., a $J_1-J_2$ Heisenberg model~\cite{Scheie2021Triangle}. The authors also found critical scaling in the dynamical structure factor near the corner of the Brillouin zone, suggesting the close proximity of this material to a second-order quantum phase transition. 

Despite a plethora of experimental studies in triangular lattice compounds, the presence and nature of a QSL phase is still under debate, as smoking-gun signals for such phases are challenging to identify. One main signature is a lack of long-range order, which is also present in other phases such as spin-glass states~\cite{Edwards1975, Rieger_book}. In fact, the actively studied spin liquid candidate YbMgGaO$_4$ has been conjectured to be a spin-glass, based on susceptibility measurements in its sister compound YbZnGaO$_4$ \cite{PhysRevLett.120.087201}. Another key signature is the presence of fractionalized quasi-particles which are hard to detect directly. Recent proposals to look at the entanglement content of the triangular lattice compound KYbSe$_2$ \cite{Scheie2021Triangle}, through the quantum Fisher information \cite{PhysRevA.85.022321, Hauke2016} and other entanglement measures \cite{PhysRevA.73.012110, PhysRevA.61.052306, PhysRevA.69.022304, PhysRevLett.93.167203, Baroni_2007, PhysRevA.74.022322}, may prove fruitful. This challenge calls for further theoretical understanding, and improved numerical simulations of experimentally relevant quantities to identify signatures of QSL phases.

Neutron scattering is potentially an excellent experimental tool to detect quantum spin liquid physics, as it directly probes the excitations in the system through the spin-spin correlation function \cite{VanHove1954, Sturm1993}. On the theoretical side, making a direct comparison with neutron scattering experiments requires calculating the dynamical structure factor, which is notoriously difficult to compute. QMC struggles to probe this quantity directly, and relies on analytic continuation from imaginary time simulations~\cite{Jarrell1996}. However, analytic continuation is numerically ill-posed due to the inherent statistical uncertainty of Monte Carlo sampling. Nevertheless, QMC supplemented by the maximum entropy method~\cite{PhysRevB.41.2380,PhysRevB.44.6011} for analytic continuation or the stochastic analytic continuation~\cite{shao2022} is still the dominant method to probe spectral functions in two- and higher-dimensional systems, with reliable results obtained in various frustration-free contexts~\cite{PhysRevLett.118.147207,Shao2017,PhysRevB.98.094403,PhysRevB.97.104424,PhysRevB.100.094411,PhysRevB.103.024403}. Unfortunately, frustrated systems, such as the triangular lattice Heisenberg model, plague QMC with the infamous sign problem, preventing efficient simulations~\cite{Loh1990,PhysRevLett.94.170201}.

In one dimension, DMRG \cite{White1992}, and the later reformulation in terms of matrix product states (MPS) \cite{Schollwock2011}, have been revolutionary. Their main success is due to the entanglement area-law in gapped systems, which leads to a finite entanglement entropy even in the thermodynamic limit~\cite{Hastings_2007}. Even for gapless one-dimensional systems, the deviations from the exact answer are understood through a finite-entanglement scaling analysis~\cite{Pollmann2009}. However, in two dimensions, an area law state still has an entanglement entropy that grows with the system size, which makes standard MPS calculations struggle to capture the thermodynamic limit. Other tensor-network-based approaches, such as projected entangled pair states (PEPS)~\cite{Verstraete2004, Cirac2019, Cirac2021}, have been proposed to work in higher dimensions. Recent work using PEPS to study the dynamical structure factor in a model near a QSL phase~\cite{Chi2022}, has found great accuracy in comparison with neutron scattering experiments in Ba$_3$CoSb$_2$O$_9$ \cite{Macdougal2020}.

Utilizing DMRG on the triangular lattice wrapped into a cylinder, yielding a quasi-one-dimensional system, has proven useful in studying static properties of QSL states~\cite{PhysRevB.92.041105, PhysRevB.92.140403, PhysRevB.94.121111, Hu_2019, Gong_2019, Szasz2020, Aghaei_2020_arxiv, Cookmeyer_2021, Szasz_2021_PRB, Jiang2021, Jiang_2022_arxiv}. We focus on this approach in this study, and extend this work into the realm of dynamics, with a similar method as was used in Ref. \onlinecite{Verresen2019}. We use this approach to examine the full phase diagram of the $J_1-J_2$ Heisenberg model on the triangular lattice.

The outline for the paper is the following. In Sec. \ref{sec:methods} we define the model we study, the quantities we examine, and the method we use to compute the dynamical spin structure factor. Next, in Sec. \ref{sec:QSL_signatures}, we discuss the three dominant proposed spin liquid states of the $J_1-J_2$ Heisenberg model, namely the gapped $\mathbb{Z}_2$, gapless $U(1)$ Dirac, and the spinon Fermi surface quantum spin liquids states. Furthermore, we outline the distinct features of these three phases to look for in the dynamical structure factor. 

In the results section, Sec. \ref{sec:results}, we first look at the nearest-neighbor Heisenberg model on the square lattice as a benchmark for our simulations. We compare with state-of-the-art QMC simulations~\cite{Shao2017, PhysRevB.92.195145} and linear spin wave theory (SWT) \cite{PhysRevB.72.014403}. The excellent agreement both qualitatively and quantitatively, justifies our method for exploring the triangular lattice in Sec. \ref{sec:results_triangular}. We first look at the 120$^{\circ}$ magnetic ordered phase, and compare our results with linear SWT~\cite{PhysRevB.79.144416}, prior numerical simulations \cite{Verresen2019, Ferrari2019, Chi2022}, Schwinger-boson theory~\cite{Ghioldi_2022_SBT}, as well as neutron scattering in Ba$_3$CoSb$_2$O$_9$ \cite{Ito2017,Macdougal2020} and KYbSe$_2$~\cite{Scheie2021Triangle}. We then look at the stripe ordered phase as a reference.

Lastly, in Sec. \ref{sec:results_qsl}, we examine the QSL phase, as well as the dependence of the spectrum on the next-nearest neighbor coupling $J_2$ through the entire phase diagram. We first look at the spectrum deep in the QSL phase, and find a V-shaped spectrum near the $\Gamma$ point, a key signature of the spinon Fermi surface state \cite{PhysRevB.96.075105}, that has been observed in NaYbSe$_2$ \cite{Dai_2021}, and YbMgGaO$_4$ \cite{Shen2016, Shen2018}. However, we also find at low energies that a gap opens near the $\Gamma$ point, which rules out such a state, as the spectrum should be gapless across the full Brillouin zone \cite{Savary2016}. Although not much is known about the microscopic model of these materials beyond them containing spin-half degrees of freedom on a triangular lattice, this discrepancy either means that this gap is at an energy scale below what was accessible in these experiments, or that these materials are not fully captured by the $J_1-J_2$ Heisenberg model. We also find that the spectrum at $\boldsymbol{q}=K$ remains gapless from the 120$^{\circ}$ phase into the QSL phase, in agreement with Schwinger-boson theory \cite{Ghioldi_2022_SBT, Scheie2021Triangle}. In contrast, we find that the gap at $\boldsymbol{q}=M$ closes as the quantum critical point is approached, ruling out the gapped $\mathbb{Z}_2$ spin liquid state. We find isolated gapless excitations at $\boldsymbol{q}=K$ and $\boldsymbol{q}=M$ throughout the entire QSL phase, in agreement with prior variational QMC~\cite{Ferrari2019} and large scale DMRG results \cite{Hu_2019}. These results strongly suggest that the QSL phase of the $J_1-J_2$ Heisenberg model is described by a gapless $U(1)$ Dirac spin liquid, and they provide clear signatures that can be detected in future neutron scattering experiments. Lastly, in Sec. \ref{sec:conclusions}, we conclude with a summary of our results, and perspectives for future studies.

\section{Models, Definitions, and Methods} \label{sec:methods}

\begin{figure}[t]
    \centering
    \quad \includegraphics[width=\linewidth]{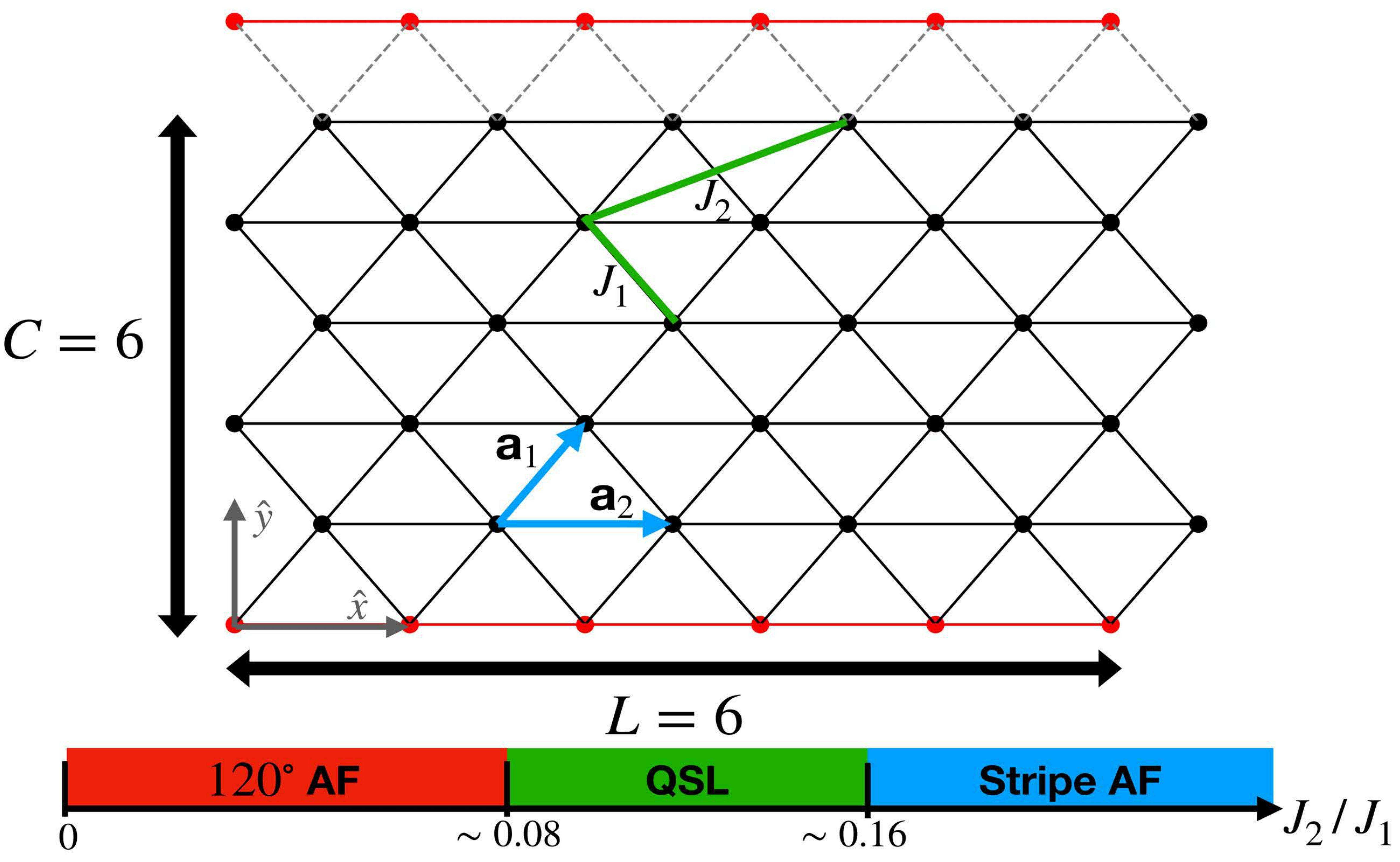}
    \caption{A $6\times 6$ triangular lattice illustrating the relevant parameters used in this work. An example of a nearest neighbor couplings $J_1$ and next-nearest neighbor couplings $J_2$ are shown in green. We also show the circumference $C$, length $L$. The lattice vectors $\boldsymbol{a_1}=(1/2,\sqrt{3}/2)$ and $\boldsymbol{a_2}=(1,0)$ are shown in blue. We also show the three expected phases of the Hamiltonian given in Eq.~\eqref{eq:H_J1-J2} and the approximate phase boundaries \cite{Iqbal2016}. }
    \label{fig:triangular_lattice}
\end{figure}

In this work, we primarily focus on the spin one-half Heisenberg model with nearest and next-nearest neighboring interactions on the triangular lattice,
\begin{equation}
    H = J_1\sum_{\langle{i,j}\rangle}\boldsymbol{S}_{i}\cdot\boldsymbol{S}_j + J_2\sum_{\langle\hspace{-2pt}\langle{i,j}\rangle\hspace{-2pt}\rangle}\boldsymbol{S}_{i}\cdot\boldsymbol{S}_j,
    \label{eq:H_J1-J2}
\end{equation}

where $\boldsymbol{S}_i=(S^x_i, S^y_i, S^z_i)$ are spin-$1/2$ operators; $\langle{i,j}\rangle$  and $\langle\hspace{-2pt}\langle{i,j}\rangle\hspace{-2pt}\rangle$ denote nearest- and next-nearest neighbor exchange interactions, respectively. We show in Fig.~\ref{fig:triangular_lattice} the lattice with circumference $C$ (we will use periodic boundary conditions along this direction in the following), length $L$, lattice vectors $\boldsymbol{a_1}$ and $\boldsymbol{a_2}$, the couplings $J_1$ and $J_2$, and the three expected phases of this model~\cite{PhysRevB.92.041105,PhysRevB.92.140403,Iqbal2016,PhysRevB.94.121111,PhysRevB.95.035141,PhysRevB.96.075116,Hu_2019}. We also examine the same Hamiltonian on the square lattice with only nearest-neighbor interactions ($J_2=0$, Heisenberg model) which serves as a benchmark to compare our results against quantum Monte Carlo supplemented by analytic continuation following Ref.~\onlinecite{Shao2017}. We set $\hbar=J_1=1$ in the following.

\subsection{Spectral Function}

The spectral function of interest in this study is the dynamical spin structure factor, relevant for neutron scattering experiments~\cite{VanHove1954,Sturm1993}, defined by
\begin{equation} \label{eq:Sqw}
    S\bigl(\boldsymbol{q},\omega\bigr) = \frac{1}{N}\sum_{\boldsymbol{x}}\int_{0}^{+\infty}\frac{\mathrm{d}t}{2\pi}e^{i\left(\omega t - \boldsymbol{q}\cdot \boldsymbol{x}\right)}G\bigl(\boldsymbol{x},t\bigr),
\end{equation}
with $N$ the number of sites in the lattice.  $G(\boldsymbol{x},t)$ is a two-point spin-spin correlation function defined by
\begin{equation}
    G(\boldsymbol{x},t) = \bra{\Omega} \boldsymbol{S}_{\boldsymbol{x}}(t) \cdot \boldsymbol{S}_{\boldsymbol{c}}(0) \ket{\Omega},
\end{equation}
with $\boldsymbol{c}$ being the center site in the lattice taken to be the origin when defining $\boldsymbol{x}$, and $\ket{\Omega}$ the ground state of the Hamiltonian $H$ with energy $E_0$.

Another useful related quantity we examine is the static spin structure factor defined by
\begin{equation} \label{eq:Sq}
    S\bigl(\boldsymbol{q}\bigr) = \frac{1}{N}\sum_{\boldsymbol{x}}\cos(\boldsymbol{q}\cdot\boldsymbol{x})G\bigl(\boldsymbol{x},t=0\bigr).
\end{equation}
Since the Hamiltonian we examine is rotationally invariant and the ground state on a finite system cannot spontaneously break the continuous SU$(2)$ symmetry of the model~\eqref{eq:H_J1-J2}, it suffices to just consider the $z$-component of the spin. Hence, what we compute in this study reduces to
\begin{equation}
    G\bigl(\boldsymbol{x},t\bigr) = 3\bra{\Omega}S^z_{\boldsymbol{x}}(t)S^z_{\boldsymbol{c}}\ket{\Omega}.
\end{equation}
We drop the factor of 3 in this work. There are methods to compute the frequency dependence of Eq. \eqref{eq:Sqw} directly, such as the correction vector method~\cite{Nocera2016}, and the Chebyshev expansion method~\cite{Holzner2011,Wolf2015}. However, both of these methods require many operator-state products between some initial state and the Hamiltonian $H$. In two dimensions, the bond dimension of the matrix product operator (MPO) representation of $H$ is large, making such operations quite inefficient.

Instead, we compute the correlation functions directly by writing,
\begin{align}
    G(\boldsymbol{x},t) &= \bra{\Omega}e^{iHt}S^z_{\boldsymbol{x}}e^{-iHt}S^z_c\ket{\Omega}\nonumber\\
    &= e^{iE_0t}\bra{\Omega}S^z_{\boldsymbol{x}}e^{-iHt}S^z_c\ket{\Omega}.
\end{align}

Then the computation of $G(\boldsymbol{x},t)$ is reduced to finding the ground state $\ket{\Omega}$, time evolving the state $S^z_c\ket{\Omega}$, and then computing its matrix elements of $S^z_{\boldsymbol{x}}$ with the ground state for all positions $\boldsymbol{x}$. 

In the definition of $S(\boldsymbol{q},\omega)$ in Eq.~\eqref{eq:Sqw}, formally infinite time and infinite space data is required, but this is not possible numerically. This forces us to truncate at a maximum distance $\boldsymbol{R}_\mathrm{max}$ and time $T_\mathrm{max}$. Introducing such a cutoff is not unique, and we discuss practical advice on how to extract the dynamical structure factor from only finite data in the following.

\begin{figure}[t]
    \centering
    \includegraphics[width=0.95\linewidth, height=10cm]{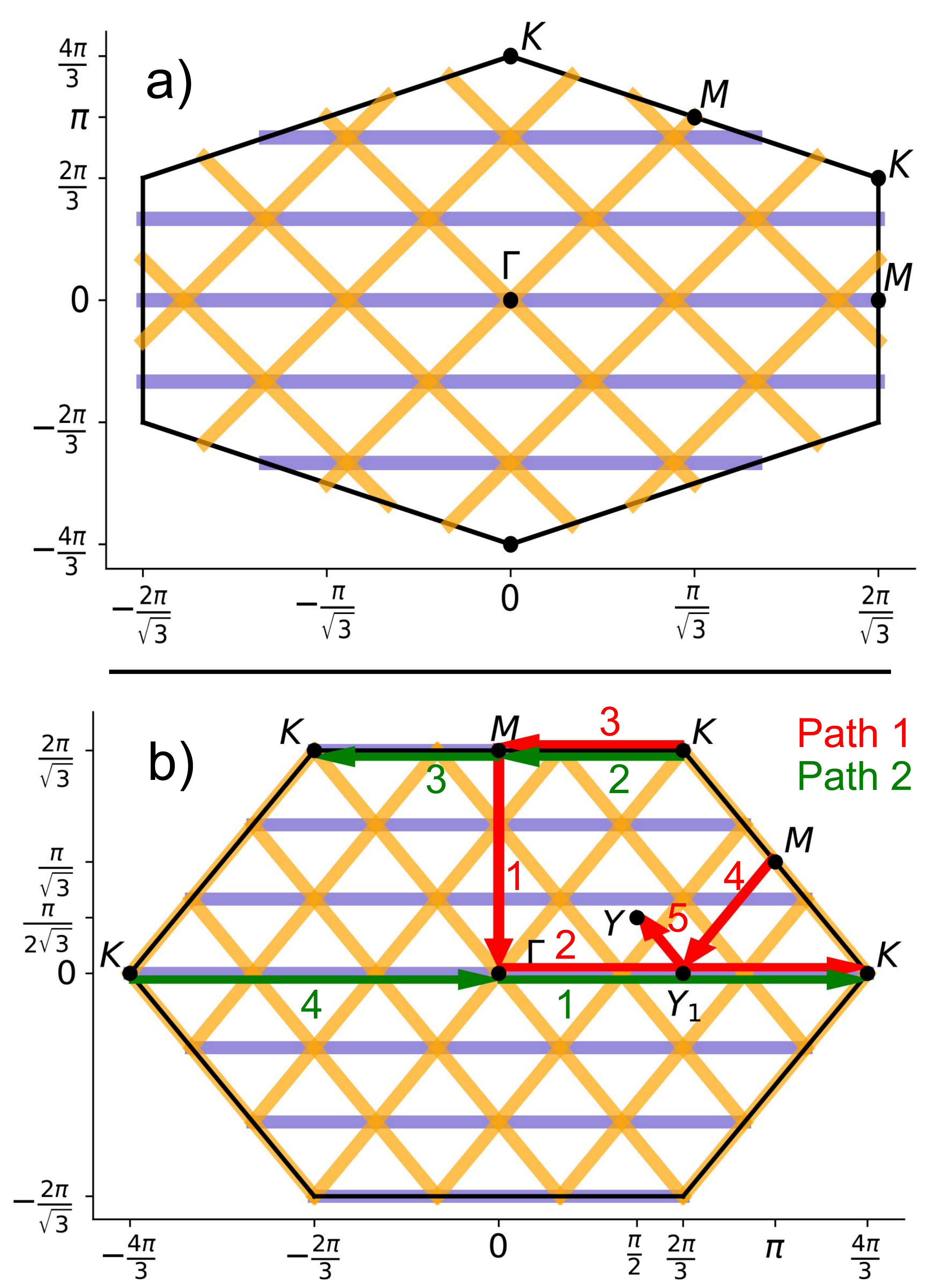}
    \caption{Allowed $\boldsymbol{q}$ values highlighted in blue for the triangular lattice wrapped into a cylinder with a circumference $C=6$. We form the cylinder using $YC$ boundary conditions in a) and $XC$ boundary conditions in b). The orange lines in both figures are the $\boldsymbol{q}$ values that are rotations of the allowed $\boldsymbol{q}$ values by the $C_6$ symmetry of the triangular lattice. In green and red are the two paths through the Brillouin zone that we examine. We note that $Y=M/2$ and $Y_1=K/2$.}
    \label{fig:triangular_q_BZ}
\end{figure}

\subsection{Finite Geometry}
\label{sec:finite-geometry}

How to approximately represent the infinite system with a finite one is not unique, and each representation has its own pros and cons. Possibly the first choice to consider is just a finite patch of the infinite system, which we will call open boundary conditions. When using MPS, we need to represent the finite system as a quasi-one-dimensional system, and doing this creates long-range interactions. If the two-dimensional system has a depth $C$ and a length $L\geq C$, then the long-range interactions are at best $O(C)$. If we use the the standard Schur form to represent $H$ as an MPO \cite{Michel2010}, then the bond dimension will also be $O(C)$. If we roll the lattice into a cylinder with circumference $C$, this causes a minimal decrease in computational efficiency as compared to open boundary conditions. Since this partially restores the translation symmetry of the infinite system with a marginal penalty, this is considered standard practice. Due to computational limitations, $C$ is quite small, and so we take $L \gg C$.

Since $C$ is small, this restricts the allowed $\boldsymbol{q}$ values quite dramatically. How one forms a cylinder out of the triangular lattice is subtle, as the choice of boundary conditions modifies which momenta in the Brillouin zone are allowed. In the literature, there are two primary boundary conditions for the triangular lattice denoted as the $XC$ and $YC$ geometries \cite{Szasz2020}. The boundary conditions in Fig. \ref{fig:triangular_lattice} is the $XC$ boundary condition, and the $YC$ boundary condition would be if we identified the left and right edges rather than the top and bottom. In Ref.~\onlinecite{Szasz2020}, the authors recommend the $YC$ boundary conditions generically, as the circumference is larger in units of the lattice spacing for the same number of lattice points. However, by examining Fig. \ref{fig:triangular_q_BZ}, we see that the $\boldsymbol{q}$ values permitted by these two geometries are dramatically different. Since most of the high symmetry $\boldsymbol{q}$ values are permitted by the $XC$ geometry, we use the $XC$ geometry throughout this work.

To find the allowed $\boldsymbol{q}=(q_x,q_y)$ values for the $XC$ geometry, we note that two conditions need to be met due to the periodic boundary conditions along the circumference and open boundary conditions along the length. For the periodic boundary conditions, we need
\begin{align}
    e^{i\boldsymbol{q}\cdot C \boldsymbol{a_1}} &= e^{i\boldsymbol{q}\cdot \frac{C}{2} \boldsymbol{a_2}} \implies e^{i\frac{\sqrt{3}}{2}C q_y} = e^{i 2\pi n} \\
    q_y &= \frac{4\pi}{\sqrt{3} C} n,\,\,\,\, n \in \left(-\frac{C}{2} , \frac{C}{2}\right]\cap \mathbb{Z}
\end{align}
The allowed values for $q_x$ are the standard allowed values for a system with length $L$, meaning
\begin{equation}
    q_x = \frac{2\pi}{L}n,\quad n\in\left(-\frac{L}{2}, \frac{L}{2}\right]\cap \mathbb{Z}.
\end{equation}

We can improve upon the heavily restricted allowed $\boldsymbol{q}$ values by generating other points in the Brillouin zone by rotating the $\boldsymbol{q}$ value by a symmetry in the lattice, as discussed in Ref.~\onlinecite{Verresen2019}. In particular, if $R$ is a rotation that leaves the lattice invariant, then in the thermodynamic limit any momentum resolved operator $O$ satisfies
\begin{equation}
    O(R\boldsymbol{q}) = O(\boldsymbol{q}).
\end{equation}
We use this relation when examining quantities on the entire Brillouin zone, to fill in much of the Brillouin zone to gain a better glimpse into the thermodynamic result. We show the allowed $\boldsymbol{q}$ values in blue, and the additional $\boldsymbol{q}$ values generated in this manner in orange in Fig. \ref{fig:triangular_q_BZ}.

Other geometries are possible, but we have found these choices the most relevant. Other possible procedures to form a cylinder out of the infinite plane are discussed in Ref.~\onlinecite{Szasz2020}.

\subsection{Time Evolution}

The standard MPS time evolution procedure in one dimension is the Time-Evolving Block Decimation (TEBD) method~\cite{Vidal2004}. This method expresses the time evolution operator in terms of unitary gates acting only on the bonds in the model. This method is exceptionally well-suited for models with only nearest neighbor interactions, but the quasi-one-dimensional systems we study here have long-range interactions. One can implement swap gates to bring distant sites near each other, and then apply the unitary gate, but this becomes inefficient rapidly as the circumference $C$ is increased. Moreover, each time step in this method increases the bond dimension, which then requires a truncation of the resulting state, which introduces errors.

An in-depth discussion of the most common time evolution techniques for MPS can be found in Ref.~\onlinecite{Paeckel2019}. In this work, we opt to use the Time-Dependent Variational Principle (TDVP) \cite{Zauner2018, Vanderstraeten2019, Yang2020} to implement the time evolution. This method automatically finds the optimum time-evolved state at the given bond dimension, and obeys conservation laws such as energy.

For any finite system, Eq. \eqref{eq:Sqw} will be come a finite sum of delta functions, as opposed to an analytical function in the thermodynamic limit. To remedy this, we broaden the peaks by convolution with a distribution, typically a Gaussian. This also doubles to serve the role of controlling the truncation of the infinite time integral in a smooth way rather than a sharp cutoff at some max time $T_\mathrm{max}$. Formally, we write
\begin{equation}
    G(\boldsymbol{x},t) \longrightarrow f_{\eta}(t) G(\boldsymbol{x},t)
\end{equation}
Then we can choose $f_{\eta}$ to be a properly normalized dampening factor. The typical choices for dampening are
\begin{equation}
    f_{\eta}(t) = \left\{\begin{matrix}
        \Theta\left(t-\eta^{-1}\right) & \rm{Sharp} \\
        e^{-\eta^2 t^2} & \rm{Gaussian} \label{eq:broad_gauss}\\
        e^{-\eta |t|}\bigr/\pi & \rm{Lorentzian}
    \end{matrix} \right.
\end{equation}

The dampening factor labelled sharp is equivalent to truncation of the time integration in Eq. \eqref{eq:Sqw} at a maximum time $T_\mathrm{max} = \eta^{-1}$. The broadening factor is an inverse time scale that is taken to be $\eta \sim O(T_\mathrm{max}^{-1})$, where $T_\mathrm{max}$ is the maximum times reliably obtained during the time evolution process. The choice of dampening factor to use depends on the problem of interest, but the general purpose choice is the Gaussian dampening. Note though, that the introduction of broadening by a Gaussian will modify the intensity and sharpness of peaks in the spectrum, and so if precision in the peaks is required, using the sharp cutoff can be useful~\cite{Scheie2021}.

We want to note that there is nothing that prevents one from time evolving to arbitrarily large times, as this just requires longer run times of the simulations. However, it is not the case that the data is necessarily reliable for these larger times. To determine the reliability of the time evolved data, we use a physically motivated criterion, namely that the spectral function must be positive for all frequencies. We broaden the spectral function using the Gaussian dampening factor in Eq. \eqref{eq:broad_gauss}, and we choose $\eta$ as small as possible so that the spectral function is positive. The maximum time $T_{\mathrm{max}}$ for which the correlation function is reliable is approximated by $T_\mathrm{max} \sim \eta^{-1}$.

\subsection{Fourier Transform}
There are two Fourier transforms necessary to achieve $S(\boldsymbol{q},\omega)$, one in space and one in time. Since we only have data for finite time and finite space, we must truncate the integrals in Eq. \eqref{eq:Sqw}. One issue with just naively truncating Eq. \eqref{eq:Sqw} is that $S(\boldsymbol{q},\omega)$ generically will acquire a non-zero imaginary part that is not physical. From there, one could only look at the real part, or the magnitude, but we propose an alternative that enforces reality of $S(\boldsymbol{q},\omega)$.

If we have a translationally invariant system, then we have the following properties,
\begin{equation}
    G(-\boldsymbol{x},t)=G(\boldsymbol{x},t)~~\mathrm{and}~~G(\boldsymbol{x},-t)=G(\boldsymbol{x},t)^*.
\end{equation}
With these properties, we can write
\begin{align}\label{eq:ft_final}
    S\bigl(\boldsymbol{q},\omega\bigr) &= \frac{1}{\pi \sqrt{N}}\int_0^{\infty}\mathrm{d}t\sum\nolimits_{\boldsymbol{x}} \cos\bigl(\boldsymbol{q}\cdot\boldsymbol{x}\bigr)  \nonumber\\ &\times\Bigl(\cos(\omega t)\mathsf{Re} G(\boldsymbol{x},t) - \sin(\omega t)\mathsf{Im} G(\boldsymbol{x},t)\Bigr).
\end{align}

\subsection{Simulation Parameters}
\begin{figure}
    \centering
    \includegraphics[width=\linewidth]{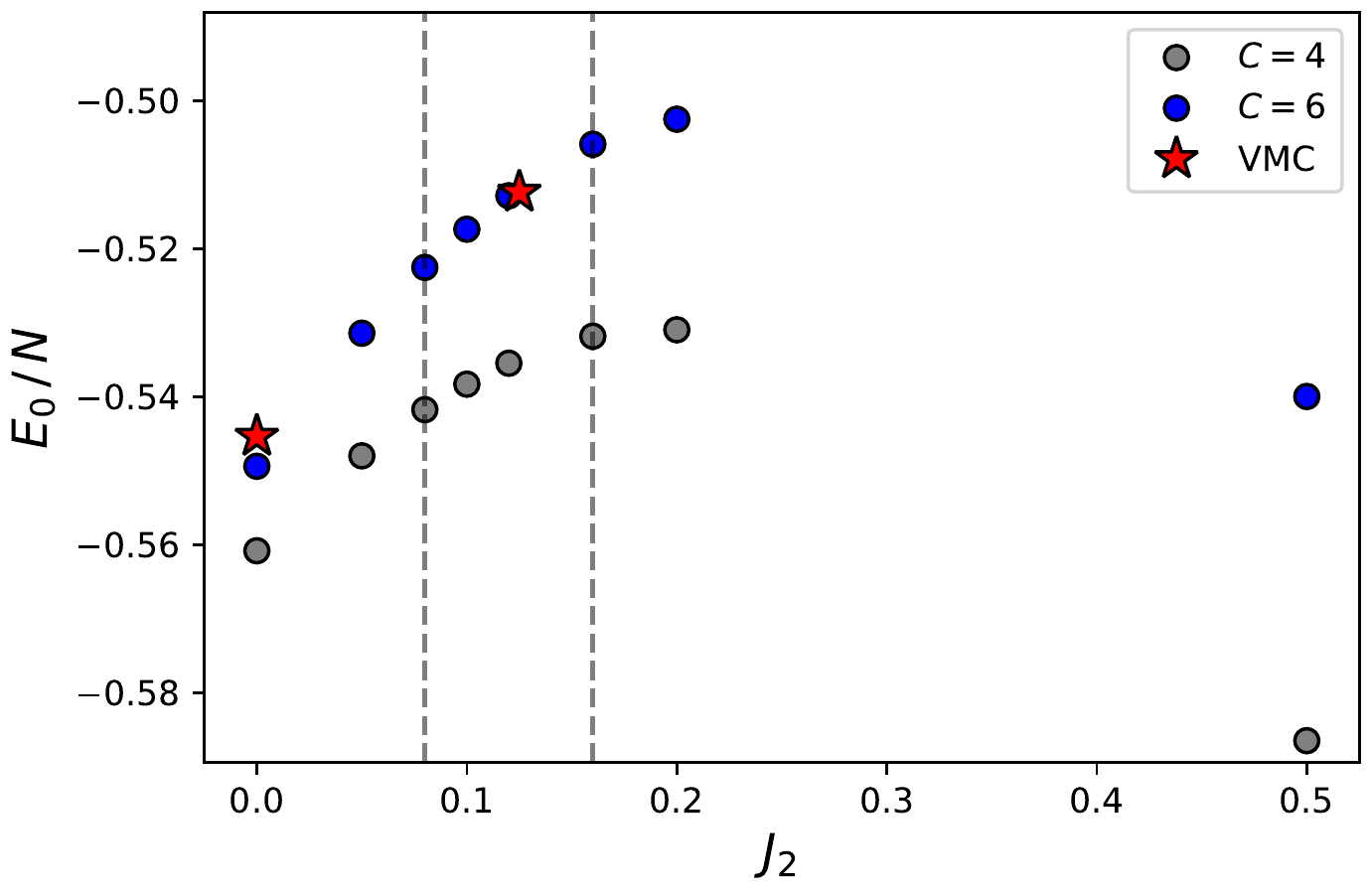}
    \caption{Ground state energy per site for the $J_1-J_2$ Heisenberg model on the triangular lattice for the $J_2$ values examined in this study. To find the ground state, we used DMRG with a bond dimension $\chi = 512$. We use a cylindrical geometry with circumference $C$ and length $L=C^2$. The stars represent the best estimate of the infinite system ground states, using variational QMC, reported in Ref. \onlinecite{Iqbal2016}.}
    \label{fig:J2_energies}
\end{figure}

In these simulations, we always work with a cylinder of circumference $C=6$ and length $L=36$. We also use a bond dimension $\chi=512$. The ground state is obtained using the DMRG~\cite{White1992, Schollwock2011} working in the zero magnetization sector with the magnetization conserved. The time evolution is performed using single-site TDVP~\cite{Zauner2018, Vanderstraeten2019, Yang2020} with a time step of $\delta t=0.1$, and a maximum time $T_\mathrm{max}=40$. The simulations are performed in real-space, and the Fourier transform to $\boldsymbol{q}$ and $\omega$ is performed using Eq.~\eqref{eq:ft_final}, with a Gaussian broadening factor as in Eq.~\eqref{eq:broad_gauss}. The broadening width $\eta$ used is $\eta^2=0.03$ for the square lattice, and $\eta^2=0.02$ for the triangular lattice in the 120$^{\circ}$ and QSL phases, and $\eta^2=0.05$ in the striped antiferromagnetic phase. For the triangular lattice, we use the $XC$ geometry as described in Sec.~\ref{sec:finite-geometry}. Our simulations utilize the ITensor library~\cite{itensor}.

In Fig. \ref{fig:J2_energies} we show the ground state energy per site as a metric for the accuracy of the ground state. We note that with only two circumferences, a finite size scaling analysis is difficult. However, we also show the best estimates for the infinite system energy density for $J_2=0$ and $J_2=0.125$ reported in Ref. \onlinecite{Iqbal2016} as a reference.

\section{Quantum Spin Liquid Signatures}\label{sec:QSL_signatures}
\begin{table}[t]
    \begin{tabular}{|c|c|}
    \hline
         Spin liquid state & Signature in $S(\boldsymbol{q},\omega)$ \\
    \hline\hline 
    \multirow{2}{6em}{Gapped $\mathbb{Z}_2$} & $\circ$ At the 120$^{\circ}$ transition, \\ 
    & gap closes at only $\boldsymbol{q}=K$.\\
    \hline
    \multirow{2}{6em}{Gapless $U(1)$ Dirac} & $\circ$ gapless at \\ 
    & $\boldsymbol{q} = K$ and $\boldsymbol{q} = M$.\\
    \hline
    \multirow{3}{6em}{Spinon Fermi surface} & $\circ$ V-shape at $\boldsymbol{q}=\Gamma$. \\ 
    & $\circ$ Broad continuum. \\
    & $\circ$ $S(\boldsymbol{q},\omega=0^+)>0\,\,\forall \boldsymbol{q}$.\\
    \hline
    \end{tabular}
    \caption{Table of the three spin liquid candidates, and their corresponding signatures in the dynamical structure factor, $S(\boldsymbol{q},\omega)$, defined by Eq. \eqref{eq:Sqw}. }
    \label{tab:QSL_signatures}
\end{table}

There are currently three dominant predictions about the nature of quantum spin liquid ground state of the $J_1-J_2$ Heisenberg model on the triangular lattice, given by Eq. \eqref{eq:H_J1-J2}.
These predictions are a gapped $\mathbb{Z}_2$ spin liquid \cite{Zhu_2015_QSL, PhysRevB.94.121111, PhysRevB.92.140403, Ghioldi_2022_SBT, Jiang_2022_arxiv}, a gapless $U(1)$ Dirac spin liquid \cite{Iqbal2016, Ferrari2019, Hu_2019}, and a spinon Fermi surface \cite{Gong_2019}. In this work, we will focus on the signatures in the dynamical structure factor, $S(\boldsymbol{q},\omega)$ in Eq. \eqref{eq:Sqw}, for these three spin liquids. For all three spin liquid states, the low-energy theory is formulated in terms of spinons, which can be understood from the parton construction \cite{Savary2016}. In this formalism, the spin operator is written as
\begin{equation}
    \boldsymbol{S} = \frac{1}{2}f_{\alpha}^{\dagger}\boldsymbol{\sigma}_{\alpha\beta}f{_\beta}, \quad f^{\dagger}_\alpha f_\alpha = 1
\end{equation}
where $f_\alpha$ are spin-1/2 fermions, $\alpha,\beta$ represent either spin up or spin down, and repeated indices are summed over. This construction has an inherent $\mathbb{Z}_2$ and $U(1)$ redundancy. Because of this, the low-energy theory can be promoted to fermions coupled to either a $\mathbb{Z}_2$ or $U(1)$ gauge field, producing the associated quantum spin liquid state. In the case of gapped $\mathbb{Z}_2$ spin liquids, the spinons can be bosons, but a gapless state is unstable to boson condensation \cite{Savary2016}. The signatures for each spin liquid state is summarized in Table \ref{tab:QSL_signatures}, and we provide an explanation for these predictions in this section.

First let us look at the $\mathbb{Z}_2$ spin liquid. In Ref. \onlinecite{PhysRevB.74.174423}, the authors discuss two possibilities for a gapped $\mathbb{Z}_2$ spin liquid on the triangular lattice, called the zero-flux and $\pi$-flux states. In the zero-flux state, the spinon dispersion relation is minimized at the corner of the Brillouin zone, $\boldsymbol{q}=K$. The magnons, which are two-spinon bound states, thus order at $\boldsymbol{q}=2K=K$. For the $\pi$-flux state, the spinon dispersion relation is minimized at $\boldsymbol{q}=Y$, and thus the magnon ordering wave-vector is $\boldsymbol{q}=2Y=M$. As the next nearest neighbor coupling is tuned, the gap closes, leading to magnon-condensation and a continuous transition into an ordered state. Via this process, the zero-flux state leads to the 120$^\circ$ state, and the $\pi$-flux state leads to the stripe ordered phase. Recent neutron scattering experiments in KYbSe$_2$ suggest that the transition from the 120$^{\circ}$ to the spin liquid phase is second order \cite{Scheie2021Triangle}. This is also observed in a prior variational QMC study \cite{Iqbal2016}, and the authors also find that the striped to spin liquid phase transition is first order. These findings are not consistent with the $\pi$-flux state. Moreover, if the ground state is the zero-flux gapped $\mathbb{Z}_2$ QSL, then a signature to look for is the spectrum being gapless at $\boldsymbol{q}=K$, and only $\boldsymbol{q}=K$, at the transition point from the 120$^{\circ}$ to the spin liquid phase. We note that precisely this prediction was observed with the Schwinger-boson formalism starting from the 120$^{\circ}$ phase \cite{Scheie2021Triangle,Ghioldi_2022_SBT}.

Next, let us examine the $U(1)$ Dirac spin liquid state. The low-energy theory on the triangular lattice is $N_f=4\,\mathrm{QED}_3$ \cite{Song2019}. In this theory, there are four spin-1/2 Dirac fermions (with two spin and two 'valley' labels), which correspond to two Dirac cones at $\boldsymbol{q}=\pm Y$ in the spinon dispersion relation. This theory permits two distinct types of gapless modes, fermion bilinears, and monopoles \cite{Song2019, PhysRevX.10.011033, PhysRevB.77.224413, PhysRevB.78.035126}.  There are in total 16 fermion bilinears \cite{PhysRevB.72.104404}, which produce gapless spin singlets and spin triplets at both $\boldsymbol{q}=\Gamma$ and $\boldsymbol{q}=M$ \cite{Song2019}. As for the monopoles, they produce spin-singlets at $\boldsymbol{q}=Y_1$, and spin-triplets at $\boldsymbol{q}=K$ \cite{Song2019, PhysRevX.10.011033}. The dynamical structure factor probes spin-1 excitations, and thus can not detect the fermion bilinears and monopoles that are spin-singlets. Also, due to the $U(1)$ symmetry in Eq. \eqref{eq:H_J1-J2}, it follows that $S(\boldsymbol{q}=\Gamma,\omega)=0$ for all $\omega$, independent of the phase. This leaves that gapless modes should be detected at $\boldsymbol{q}=K$ and $\boldsymbol{q}=M$ if the ground state is a $U(1)$ Dirac spin liquid. This signature has been observed in a prior DMRG calculation, where they looked at where the correlation length diverges under flux insertion \cite{Hu_2019}. 

The last spin liquid state to talk about is the state with a spinon Fermi surface. The idea here is that the low-energy theory is described by a metallic state with a half-filled band, leading to a Fermi surface, and many low energy excitations \cite{Shen2016, PhysRevB.96.075105}. In Ref. \onlinecite{PhysRevB.96.075105}, they start with a mean-field Hamiltonian for free fermions at half-filling of the form
\begin{align}\label{eq:H_MFT}
    H_{\mathrm{MFT}}=
    -&t_1\sum_{\langle i,j \rangle,\alpha }f_{i\alpha}^{\dagger}f_{j\alpha} 
    -t_2\sum_{\langle\hspace{-2pt}\langle{i,j}\rangle\hspace{-2pt}\rangle,\alpha }f_{i\alpha}^{\dagger}f_{j\alpha} \nonumber\\
    -&\mu \sum_{i,\alpha}f^{\dagger}_{i\alpha}f_{i\alpha}    
\end{align}
This Hamiltonian is quadratic, and thus diagonalizable, and the chemical potential enforces half-filling. Then, the dynamical structure factor is given by Eq. \eqref{eq:Sqw}, with
\begin{equation}\label{eq:G_SFS}
    G(\boldsymbol{x},t) = \bra{\Omega} S^-_{\boldsymbol{x}}(t) S^+_{\boldsymbol{c}}(0) \ket{\Omega},
\end{equation}
In terms of the fermion operators, we have
\begin{equation}
    S_{\boldsymbol{q}}^+=\sum_{\boldsymbol{k}}f^{\dagger}_{[\boldsymbol{k} + \boldsymbol{q}]\uparrow}f_{\boldsymbol{k}\downarrow}
\end{equation}

This means that $S(\boldsymbol{q},\omega=0^+) > 0$ if $\boldsymbol{q}$ can be written as $\boldsymbol{k}_1 + \boldsymbol{k}_2$, with $\boldsymbol{k}_1,\boldsymbol{k}_2$ located at the Fermi-surface. Due to the half-filling constraint, this is actually possible for every $\boldsymbol{q}$. This model makes three predictions about the dynamical structure factor: (i) $S(\boldsymbol{q},\omega=0+)>0 \,\, \forall \boldsymbol{q}$, (ii) a V-shape upper excitation edge near $\boldsymbol{q}=\Gamma$, and (iii) a broad continuum throughout the Brillouin zone, with no sharp magnon branches \cite{Shen2016, PhysRevB.96.075105, Savary2016}. Signatures for a spinon Fermi-surface have been seen in neutron scattering experiments in YbMgGaO$_4$ \cite{Shen2016, Shen2018}, as well as in NaYbSe$_2$ \cite{Dai_2021}.

\begin{figure*}[t]
    \centering
    \includegraphics[width=\linewidth, height=6.5cm]{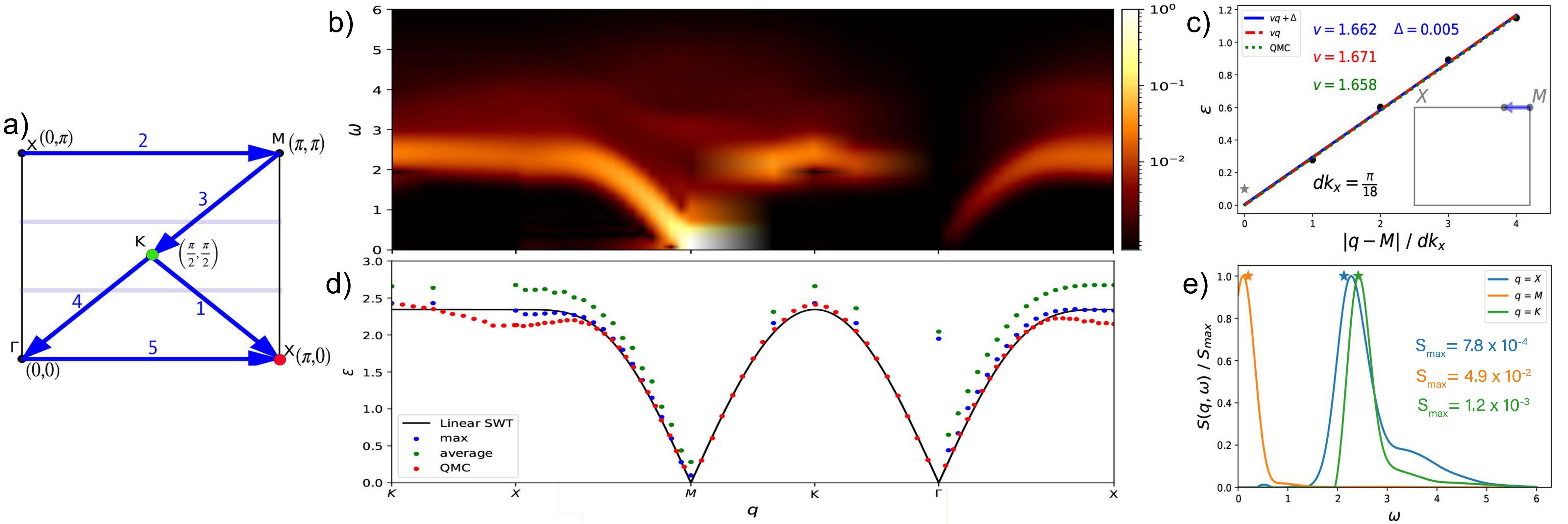}
    \caption{Results for the nearest-neighbor antiferromagnetic Heisenberg model on the square lattice, defined by Eq. \eqref{eq:H_J1-J2} with $J_2=0$. We show the dynamical structure factor $S(\boldsymbol{q},\omega)$ defined by Eq. \eqref{eq:Sqw} in b), for the path shown in a). From this, we show the magnon dispersion relation obtained using Eq. \eqref{eq:w_max} and Eq.~\eqref{eq:w_bar}, compared with linear SWT~\cite{PhysRevB.72.014403} and QMC~\cite{Shao2017} in d). The SWT result is adjusted by a common factor analogous to what is done in Ref.~\onlinecite{Shao2017}. 
    We show the dispersion relation using Eq. \eqref{eq:w_max} for momentum values near $\boldsymbol{q}=M=(\pi,\pi)$ on the path towards $\boldsymbol{q}=X=(0,\pi)$ in c). We fit these points to a line of the form $vq+\Delta$ shown in blue, also the same form but enforcing $\Delta=0$ in red, and the velocity obtained from QMC in green \cite{PhysRevB.92.195145}. We omit the value of $\epsilon(\boldsymbol{q}=M)$ in the fits. We also show the frequency dependence of $S(\boldsymbol{q},\omega)$ \eqref{eq:Sqw} at fixed high symmetry momentum values in b), and compare with the gap determined using QMC in Ref. \onlinecite{Shao2017} shown with a star. We divide the values by the maximum intensity $S_\mathrm{max}$ to view all three points on the same axis.}
    \label{fig:square_all}
\end{figure*}

\section{Results}\label{sec:results}

We examine the square and triangular lattice geometries. For the square lattice, we only look at the nearest neighbor model ($J_2=0$), which serves as a benchmark to compare this method against quantum Monte Carlo (QMC) calculations \cite{Shao2017, PhysRevB.92.195145}, and linear spin wave theory (SWT)~\cite{PhysRevB.72.014403}. QMC is not efficient on the triangular lattice due to frustration leading to the infamous sign problem~\cite{Loh1990,PhysRevLett.94.170201}, but the method of the present work has been used for a direct comparison with the neutron scattering experiments in KYbSe${}_2$ with considerable success in Ref.~\onlinecite{Scheie2021Triangle}. We provide a significantly more thorough exploration and analysis of the triangular lattice here.

\subsection{Spin-$1/2$ Heisenberg Model on the Square Lattice}

The square lattice Heisenberg model has been well-studied, and so this provides an excellent benchmark to test two-dimensional MPS time evolution. In Fig. \ref{fig:square_all} b) we show our results for the dynamical structure factor for the path taken shown in Fig. \ref{fig:square_all} a). We see that for the third and fourth segments, from $\boldsymbol{q}=M\rightarrow \Gamma$, there is a low number of allowed $\boldsymbol{q}$ values due to the small circumference of the cylinder used in the MPS simulations. We include the point $\boldsymbol{q}=K$ even though it is not formally permitted by this geometry, due to its significance. Where there are sufficient allowed momentum values, we see good qualitative agreement with the spectrum obtained using QMC in Ref. \onlinecite{Shao2017}.

The qualitative agreement is useful on its own, but we also explore the quantitative accuracy of this method. In Fig. \ref{fig:square_all} d), we show the magnon dispersion relation extracted from the spectrum. We define two methods to extract the dispersion relation, defined by
\begin{align}
    \epsilon(\boldsymbol{q}) &:= 
    \argmax\limits_{\omega}\,S(\boldsymbol{q},\omega) \label{eq:w_max}\\
    \langle \omega \rangle (\boldsymbol{q}) &:= \frac{\int_{-\infty}^{+\infty}\mathrm{d}\omega \, \omega S(\boldsymbol{q},\omega) }{ \int_{-\infty}^{+\infty}\mathrm{d}\omega\, S(\boldsymbol{q},\omega)} \label{eq:w_bar}
\end{align}
From the comparison with SWT~\cite{PhysRevB.72.014403} and QMC in Fig. \ref{fig:square_all} d), we see that using Eq. \eqref{eq:w_max} yields a more accurate method of obtaining the dispersion relation, which is consistent with prior work~\cite{PhysRevB.98.094403}. We will focus our attention on Eq. \eqref{eq:w_max} when discussing the dispersion relation in this work. We note the value at $\boldsymbol{q}=\Gamma$ is not accurate, as there is essentially zero weight at this momentum value, and so the maximum is largely measuring machine precision rather than something physical.

From the dispersion relation, we can extract the magnon velocity $v$ near $\boldsymbol{q}=M$ by defining $\boldsymbol{v} = \partial_{\boldsymbol{q}}\epsilon(\boldsymbol{q})$. In Fig. \ref{fig:square_all} c), we show the dispersion relation near $\boldsymbol{q}=M$, and the extracted velocity. The magnitude of the velocity should be independent of the direction, so we take the path from $\boldsymbol{q}=M\rightarrow X$ as we have the most allowed $\boldsymbol{q}$ values in that direction. We exclude the value exactly at $\boldsymbol{q}=M$, as we anticipate this value to have the largest finite size effects. SWT predicts the spectrum is gapless at $\boldsymbol{q}=M$, but any finite system will always have a gap.

We fit the MPS data to a line of the form $vq + \Delta$, first allowing $\Delta$, which corresponds to the value at $\boldsymbol{q}=M$, to be a fitting parameter, as well as forcing $\Delta=0$. The square lattice Heisenberg model velocity has also been calculated using QMC \cite{PhysRevB.92.195145}. In this work, the authors determine the velocity from a hydro-dynamical relation between the velocity and static quantities, as well as a winding number based approach. These approaches do not require analytic continuation, making them quite accurate. The close agreement with our simulations is a non-trivial justification for our method. When allowing $\Delta$ to be a free fitting parameter, we see the velocity is closer to that of the velocity obtained with QMC. We note that this fit provides an estimate of the gap at $\boldsymbol{q}=M$, which we see to be around $\Delta \approx 0.005$. We also note that another measure of the gap would be just the value of the dispersion relation $\epsilon(\boldsymbol{q}=M)$, which gives a gap $\Delta \approx 0.1$. We do not anticipate that generally our methods will be a reliable means to accurately extract the gap in the thermodynamic limit. First, the finite system size produces a gap even for gapless systems, which for an $SU(2)$ invariant system we expect to scale as $\Delta \sim 1/LC$. Second, the finite time $T_\mathrm{max}$ we can achieve with this method produces effectively a minimum frequency $\omega_\mathrm{min}$ that we can resolve on the order of $\omega_\mathrm{min}\sim T_\mathrm{max}^{-1}$. We also have $T_\mathrm{max}^{-1} \sim \eta$, and so frequencies $\omega \lesssim \eta = 0.16$, are uncertain. However, this can provide an upper bound estimate to the gap.

Lastly we show the frequency dependence of the spectrum at the high symmetry $\boldsymbol{q}$ points in Fig. \ref{fig:square_all} e). We also show with a star the value for the dispersion relation from QMC in Ref. \onlinecite{Shao2017}. We see that the dispersion relation values match well with the maximums of these plots, which provides justification for Eq. \eqref{eq:w_max} as a good definition for the magnon dispersion relation.

In this section, we found our method yields a high level of agreement with QMC simulations for the spectral function and magnon dispersion relation~\cite{Shao2017}. We also were able to extract the velocity from the dispersion relation directly, and found excellent agreement with accurate estimations from QMC based on static properties~\cite{PhysRevB.92.195145}. This comparison provides confidence in our technique to explore the triangular lattice Heisenberg model, where QMC is plagued by the sign problem \cite{Loh1990,PhysRevLett.94.170201}.

\begin{figure*}[t]
    \centering
    \includegraphics[width=\textwidth, height=7cm]{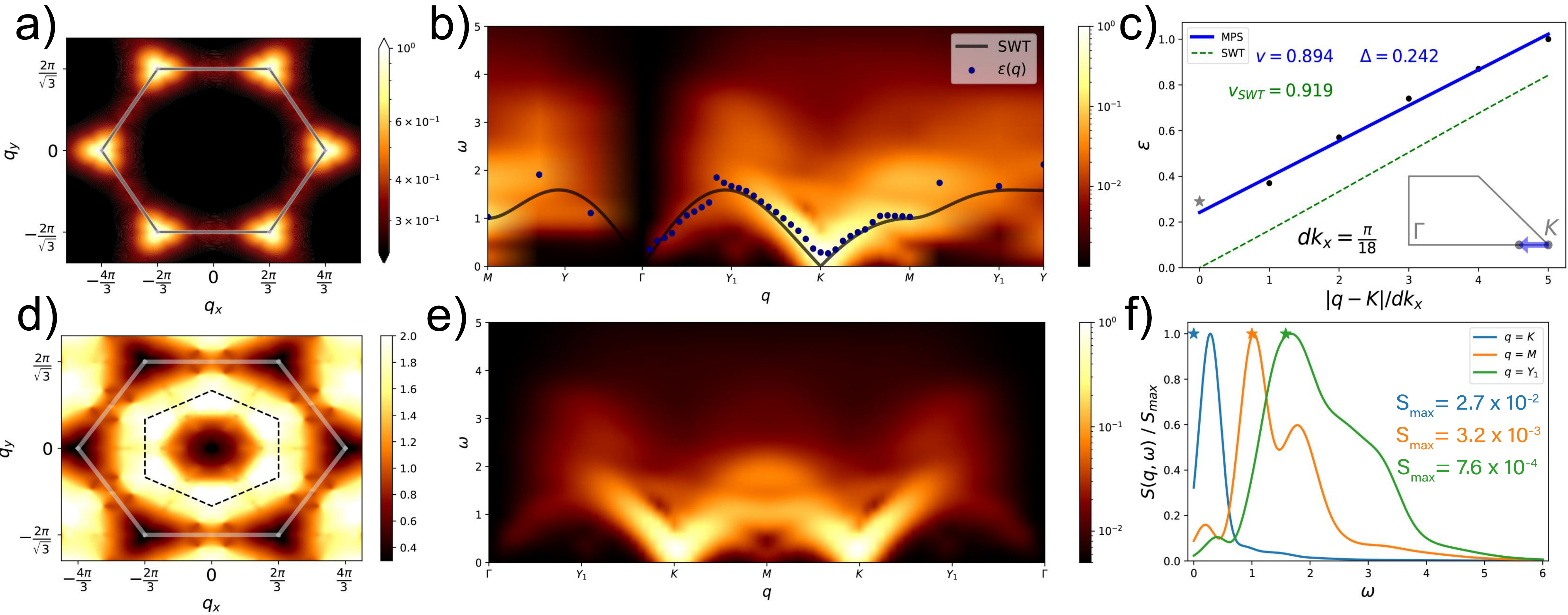}
    \caption{Results for the nearest-neighbor antiferromagnetic Heisenberg model on the triangular lattice, defined by Eq. \eqref{eq:H_J1-J2} with $J_2=0$. We show the static structure factor $S(\boldsymbol{q})$ defined by Eq. \eqref{eq:Sq} in a). In b) and e) we show the dynamical structure factor $S(\boldsymbol{q},\omega)$ defined by Eq. \eqref{eq:Sqw} for path 1 and path 2 shown in Fig. \ref{fig:triangular_q_BZ} respectively. We show the dispersion relation using Eq. \eqref{eq:w_max} for momentum values near $\boldsymbol{q}=K=(4\pi/3,0)$ on the path towards $\boldsymbol{q}=\Gamma=\boldsymbol{0}$ in c). We fit these points to a line of the form $vq+\Delta$ shown in blue, and the dispersion relation and velocity from linear SWT in green \cite{PhysRevB.79.144416}. The dispersion relation using Eq. \eqref{eq:w_max} is shown in d). Lastly, We show the frequency dependence of $S(\boldsymbol{q},\omega)$ at fixed high symmetry momentum values in f), and compare with the SWT results at those momenta shown with a star. We divide the values by the maximum intensity $S_\mathrm{max}$ to view all three points on the same axis. For both the static structure factor and the dispersion relation, we restore the 6-fold rotational symmetry of the lattice in the thermodynamic limit, as discussed in Sec.~\ref{sec:finite-geometry}.}
    \label{fig:120_all}
\end{figure*}


\subsection{Spin-$1/2$ Heisenberg $J_1-J_2$ Model on the Triangular Lattice}\label{sec:results_triangular}

The $J_1-J_2$ Heisenberg model on the triangular lattice hosts three distinct phase as we tune the coupling constant $J_2$. For intermediate values of $J_2$, there is a quantum spin liquid phase, and the exact nature of this phase is an active area of current research~\cite{PhysRevB.92.041105,PhysRevB.92.140403,Iqbal2016,PhysRevB.94.121111,PhysRevB.95.035141,PhysRevB.96.075116,Hu_2019}. Because of this, there is a great interest in reliable simulations in this region. On the experimental side, the material KYbSe$_2$ is a promising candidate to realize this QSL phase \cite{Scheie2021Triangle}. In that study, it was observed that KYbSe$_2$ is well modelled by Eq. \eqref{eq:H_J1-J2}, with $J_2 / J_1 \sim 0.05$, which is very close to boundary of the QSL phase which is approximately $J_2 / J_1 \sim 0.08$ \cite{Iqbal2016}. Applying pressure to this material could push it into the QSL phase, and so signatures of the QSL phase are desired from the theory side. We explore the full phase diagram of the $J_1-J_2$ Heisenberg model here.

\subsubsection{The $120^{\circ}$ Magnetic Long-Range Ordered Phase}

The $J_2=0$ phase of the triangular lattice Heisenberg model has a rich history starting back with Anderson postulating the ground state as a candidate for a resonating valence bond state \cite{Anderson1973}, but more recently evidence suggests the state realizes a 120$^{\circ}$ ordering antiferromagnetic state~\cite{Huse1988, Jolicoeur1989, Singh1992, Chubukov1994, Bernu1994, Capriotti1999, Zheng2006, White2007}. In Fig. \ref{fig:120_all} a) we show the static structure factor obtained from DMRG, and see ordering at $\boldsymbol{q}=K$ as expected. In Fig. \ref{fig:120_all} b), we show the full spectrum, the dispersion relation obtained using Eq. \eqref{eq:w_max}, as well as the dispersion relation from SWT~\cite{PhysRevB.79.144416}, for path 1 in Fig. \ref{fig:triangular_q_BZ}. The discontinuity in $\epsilon(\boldsymbol{q})$ in the middle of the path from $\boldsymbol{q}=\Gamma \rightarrow K$ we believe to be an artifact of the definition in Eq. \eqref{eq:w_max}, as there are two branches of near equal spectral weight in this momentum range. We also note that near $\boldsymbol{q}=K$, we anticipate $\epsilon(\boldsymbol{q})$ to overestimate the real dispersion relation due to finite size effects, and reduced resolution for small $\omega$ due to finite time.

\begin{figure*}[t]
    \centering
    \includegraphics[width=\textwidth, height=7cm]{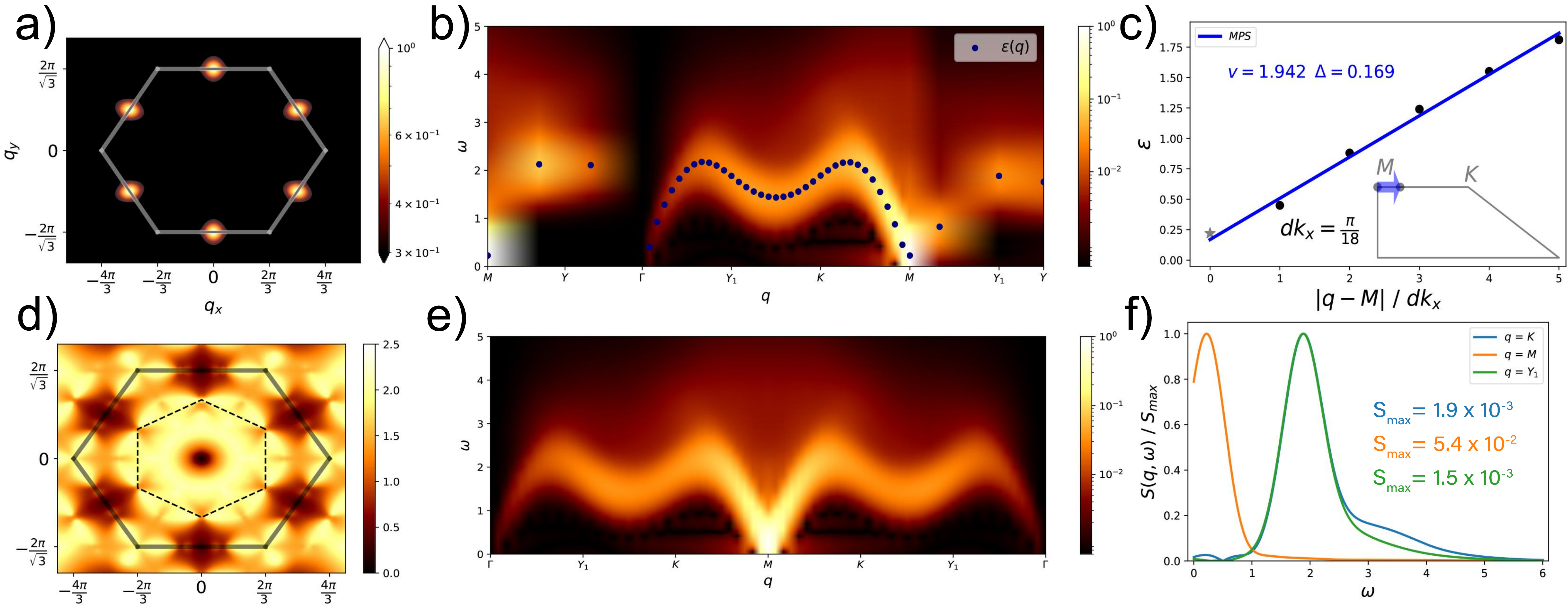}
    \caption{Results for the $J_1-J_2$ Heisenberg model on the triangular lattice, defined by Eq. \eqref{eq:H_J1-J2} with $J_2/J_1=0.5$. We show the static structure factor $S(\boldsymbol{q})$ defined by Eq. \eqref{eq:Sq} in a). In b) and e) we show the dynamical structure factor $S(\boldsymbol{q},\omega)$ defined by Eq. \eqref{eq:Sqw} for path 1 and path 2 shown in Fig. \ref{fig:triangular_q_BZ} respectively. We show the dispersion relation using Eq. \eqref{eq:w_max} for momentum values near $\boldsymbol{q}=M=(0,{2\pi}/{\sqrt{3}})$ on the path towards $\boldsymbol{q}=K=({2\pi}/{3},{2\pi}/{\sqrt{3}})$ in c). We fit these points to a line of the form $vq+\Delta$ shown in blue. The dispersion relation using Eq. \eqref{eq:w_max} is shown in d). Lastly, We show the frequency dependence of $S(\boldsymbol{q},\omega)$ at fixed high symmetry momentum values in f). We divide the values by the maximum intensity $S_\mathrm{max}$ to view all three points on the same axis. For both the static structure factor and the dispersion relation, we restore the 6-fold rotational symmetry of the lattice in the thermodynamic limit, as discussed in Sec.~\ref{sec:finite-geometry}.}
    \label{fig:striped_all}
\end{figure*}

Despite the overestimation of the dispersion relation, we compute the velocity of excitations, which would be robust to a systematic overestimation of the dispersion relation. Near $\boldsymbol{q}=K$ we anticipate the velocity to be independent of the direction, and so we take the direction from $\boldsymbol{q}=K\rightarrow\Gamma$, as this has the most allowed $\boldsymbol{q}$ values in the linear region of SWT. We show the comparison of our MPS results with SWT in Fig. \ref{fig:120_all} c). We see an almost uniform increase of the dispersion relation compared to SWT. We fit the data to a line of the form $vq + \Delta$, which provides an estimate to the velocity, as well as the gap. Again we do not include the value at $\boldsymbol{q}=K$, as this is the least accurate point since we expect gapless modes from SWT. We find the velocity is close to SWT, with a slight decrease. The fit parameter $\Delta$, as well as the value of $\epsilon(\boldsymbol{q}=K)$ provide an estimate to the gap, but this should be viewed as an upper bound rather than a quantitatively accurate result. Analogously to the square lattice case, there is a finite-size gap that we expect to scale as $\Delta \sim 1/LC$, as well as minimum frequency resolution on the order of $\omega_\mathrm{min} \sim \eta \approx 0.14$, making the gap challenging to resolve accurately.

We look at the dispersion relation for the entire Brillouin zone in Fig. \ref{fig:120_all} d). It has been previously seen that there is a reduction from the linear spin wave theory results most noticeably at the $\boldsymbol{q}=M$ and $\boldsymbol{q}=Y_1$ points in the Brillouin zone \cite{Verresen2019,Ferrari2019}. This reduction in the dispersion relation at $\boldsymbol{q}=Y_1$ leads to a surprising stabilization of quasi-particles, preventing their decay \cite{Verresen2019}. The reduction is not as clear here when looking at the dispersion relation in Fig. \ref{fig:120_all} d). However, in Fig. \ref{fig:120_all} b) and e), we can clearly see a low energy roton mode near $\boldsymbol{q}=M$, and low energy spectral weight at $\boldsymbol{q}=Y_1$, as expected. These low energy peaks are also visible in Fig. \ref{fig:120_all} f). Since these low energy branches are not the maximum intensity frequency at these momentum values, our definition of the dispersion relation does not pick them out. However, if we do not damp the data with a Gaussian factor, then these low energy peaks do become the frequency with the maximum intensity.

We show the dynamical structure factor in Fig. \ref{fig:120_all} e) for path 2 in Fig. \ref{fig:triangular_q_BZ}, to easily compare with recent neutron scattering results in Ba$_3$CoSb$_2$O$_9$ \cite{Ito2017,Macdougal2020}, PEPS results \cite{Chi2022}, and Schwinger-boson theory~\cite{Ghioldi_2022_SBT}. We find the maximum spectral weight resides at $\boldsymbol{q}=K$,  and we also see the low energy roton-like mode at $\boldsymbol{q}=M$ seen in prior simulations \cite{Verresen2019, Ferrari2019, Chi2022}, as well as neutron scattering experiments in Ba$_3$CoSb$_2$O$_9$ \cite{Ito2017,Macdougal2020} and KYbSe$_2$~\cite{Scheie2021Triangle}. We note that the roton-like mode is not fully captured in the Schwinger-boson formalism, but the behavior near $\boldsymbol{q}=K$ appears to be well captured.


\subsubsection{The Stripe Ordered Phase}



For large, $J_2 / J_1$, the Hamiltonian given by Eq. \eqref{eq:H_J1-J2} exhibits a striped antiferromagnetic ground state. We use $J_2 / J_1 = 0.5$ to study this phase. We show the full spectrum given in Fig. \ref{fig:striped_all} b) and e) for the paths shown in Fig. \ref{fig:triangular_q_BZ}. We also show the dispersion relation from Eq. \eqref{eq:w_max} in Fig. \ref{fig:striped_all} d). For both the static structure factor and the dispersion relation, we restore the 6-fold rotational symmetry of the lattice in the thermodynamic limit, as discussed in Sec.~\ref{sec:finite-geometry}. The striped phase does in fact break this symmetry, and so this would illustrate the symmetric state formed as a superposition of the three symmetry broken states.

We show the velocity near $\boldsymbol{q}=M$ in Fig. \ref{fig:striped_all} c), as well as the gap determined from the linear fit. Again we omit the value of $\epsilon(\boldsymbol{q}=M)$ from the fit, and can compare the value with the value $\Delta$ obtained in the fit. Similarly, we anticipate the largest deviations to be near where we expect gapless excitations, which is at $\boldsymbol{q}=M$ in the striped phase. In the previous systems discussed earlier, we found that the dispersion relation away from these gapless points had good quantitative agreement with SWT and/or QMC, and so we anticipate the same here. We also show the frequency dependence at $\boldsymbol{q}=K,M$, and $Y_1$ in Fig. \ref{fig:striped_all} f) for reference.


\subsubsection{The Quantum Spin Liquid Phase}\label{sec:results_qsl}

\begin{figure*}[t]
    \centering
    \includegraphics[width=\linewidth, height=7.5cm]{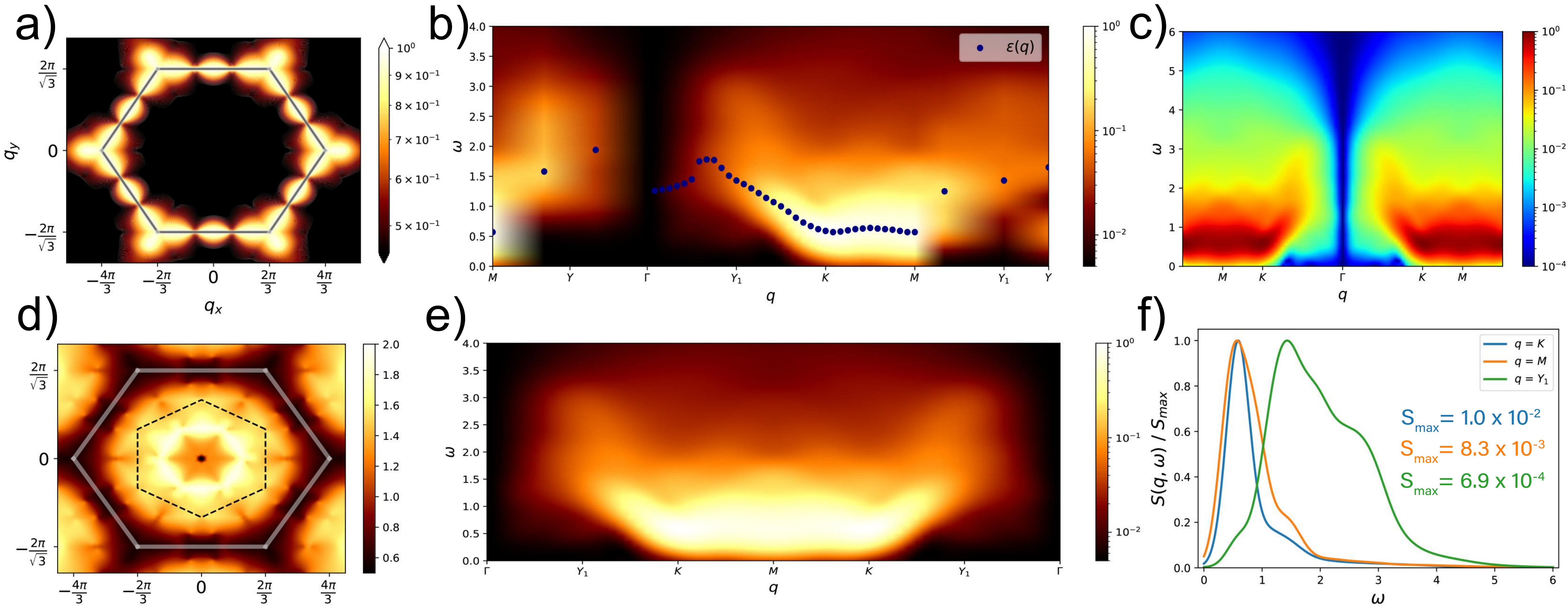}
    \caption{Results for the $J_1-J_2$ Heisenberg model on the triangular lattice, defined by Eq. \eqref{eq:H_J1-J2} with $J_2/J_1=0.12$. We show the static structure factor $S(\boldsymbol{q})$ defined by Eq. \eqref{eq:Sq} in a). In b) and e) we show the dynamical structure factor $S(\boldsymbol{q},\omega)$ defined by Eq. \eqref{eq:Sqw} for path 1 and path 2 shown in Fig. \ref{fig:triangular_q_BZ} respectively. In c) we show the spectral function using a similar path and color map as Fig. 4 in Ref. \onlinecite{Dai_2021} for easy comparison. The maximum intensity using Eq. \eqref{eq:w_max} is shown in d). Lastly, we show the frequency dependence of $S(\boldsymbol{q},\omega)$ at fixed high symmetry momentum values in f). We divide the values by the maximum intensity $S_\mathrm{max}$ to view all three points on the same axis. For both the static structure factor and the dispersion relation, we restore the 6-fold rotational symmetry of the lattice in the thermodynamic limit, as discussed in Sec.~\ref{sec:finite-geometry}.}
    \label{fig:qsl_all}
\end{figure*}

\begin{figure*}[t]
    \centering
    \includegraphics[width=\linewidth]{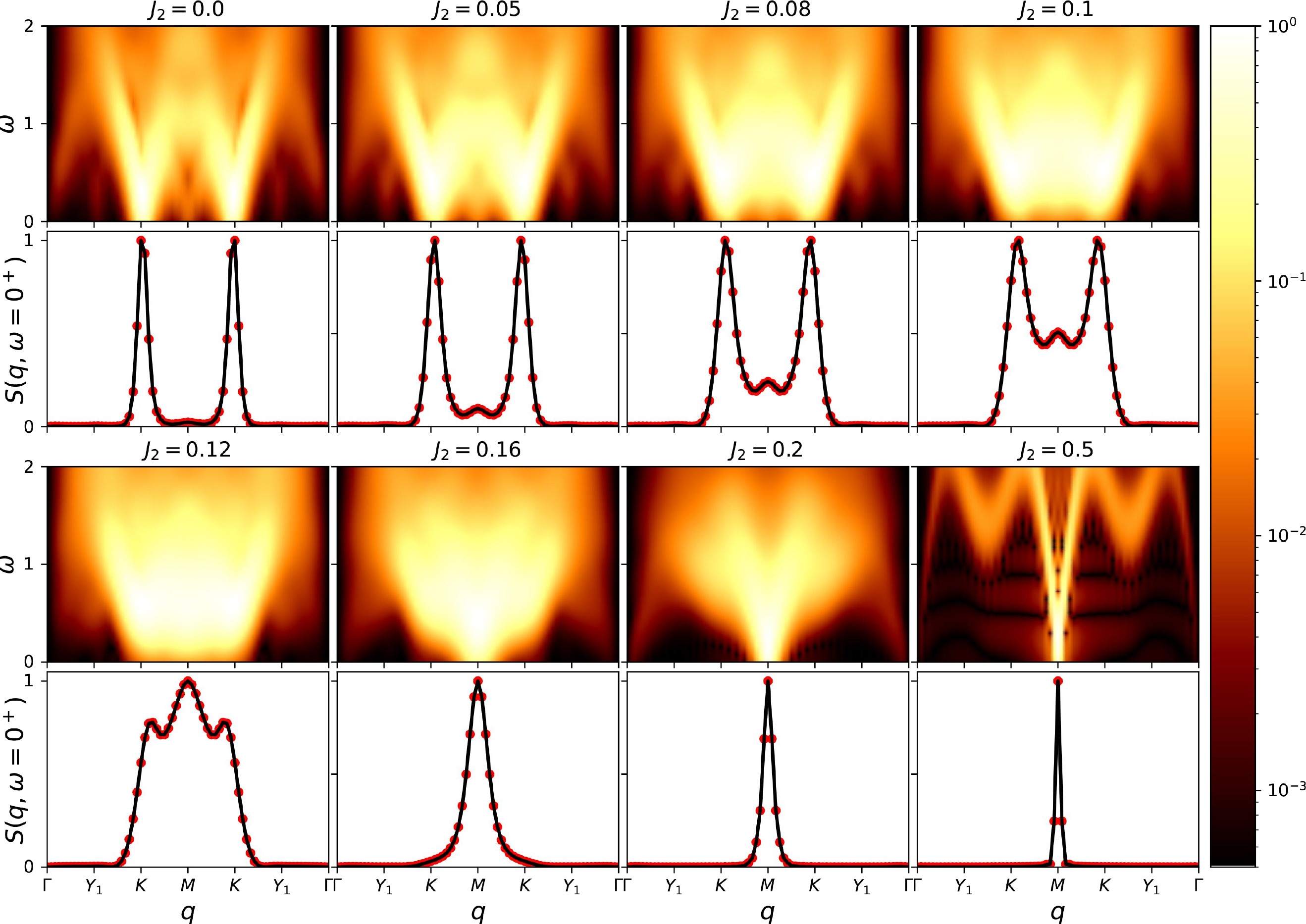}
    \caption{The dynamical structure factor given by Eq. \eqref{eq:Sqw} for the $J_1-J_2$ Heisenberg model on the triangular lattice defined by Eq. \eqref{eq:H_J1-J2} along path 2 shown in Fig. \ref{fig:triangular_q_BZ}. We show multiple $J_2$ values across the entire phase diagram. This model is in the spin liquid phase for $0.08 \lesssim J_2 \lesssim 0.16$ \cite{Iqbal2016}. We use a Gaussian broadening defined by Eq. \eqref{eq:broad_gauss}, with $\eta^2=0.02$. The first and third rows show the spectral function, and the second and last row show low frequency cuts of the spectral function. The cuts are obtained by integrating the frequency from 0 to 0.1, and then normalize so that the maximum intensity is 1. The black line is just to help with visualizing the data points in red.}
    \label{fig:J2_both}
\end{figure*}

Lastly, we examined the quantum spin liquid phase, with $J_2 / J_1 = 0.12$, which is right in the middle of the two boundaries of the QSL phase as predicted by Ref.~\onlinecite{Iqbal2016}. We find the maximum intensity is spread across the Brillouin zone boundary in the static structure factor shown in Fig. \ref{fig:qsl_all} a), a feature also present in kagome lattice spin liquid systems \cite{PhysRevLett.109.067201,PhysRevB.97.014423,Zhu_2019_kagome,PhysRevB.103.014431, Han2012}. We show the full spectrum in Fig. \ref{fig:qsl_all} b) and e), along the paths through the Brillouin zone illustrate in Fig. \ref{fig:triangular_q_BZ}. We note the low energy spectral weight along the entire Brillouin zone boundary, seen in the path from $\boldsymbol{q}=K\rightarrow M$. This is indicative of a competition between the 120$^{\circ}$ and the stripe ordered phases, which order at $\boldsymbol{q}=K$ and $\boldsymbol{q}=M$, respectively. In Fig. \ref{fig:qsl_all} b) we see a faint high-energy branch near $\boldsymbol{q}=\Gamma$, in agreement with Ref. \onlinecite{Ferrari2019}. However, we find a continuum extending to higher energies, and this branch lies fully within this continuum as opposed to being isolated. We believe this difference is because our simulations do not rely on an ansatz for the excitations, and can probe the full spectrum.

We also see the low energy spectral weight along the Brillouin zone boundary by looking at the dispersion relation shown in Fig.  \ref{fig:qsl_all} d), obtained using Eq. \eqref{eq:w_bar}. It is more accurate to call this just the maximum intensity as a function of momentum $\boldsymbol{q}$. The dispersion relation interpretation assumes well define magnon modes, which we do not expect in the QSL phase. Nevertheless, it does indicate that there is low energy spectral weight across the Brillouin zone boundary, but low energy spectral weight is absent near the center of the Brillouin zone. Lastly we show the frequency dependence of $S(\boldsymbol{q},\omega)$ at the high symmetry $\boldsymbol{q}$ points in Fig. \ref{fig:qsl_all} f). We want to emphasize the similarity between $\boldsymbol{q}=K$ and $\boldsymbol{q}=M$, which is unique to the QSL phase. We also note that the maximum intensity is suppressed as compared to the 120$^{\circ}$ phase in Fig. \ref{fig:120_all} f), and the striped phase in Fig. \ref{fig:striped_all} f). 

Let us now examine what these results have to say about the nature of the QSL ground state. First, we show in Fig. \ref{fig:qsl_all} c) the spectrum with a similar colormap and momentum path as Fig. 4 in Ref \onlinecite{Dai_2021} for easy comparison. We see the V-shape spectrum near $\boldsymbol{q}=\Gamma$, as is observed in previous neutron scattering experiments in NaYbSe$_2$ ~\cite{Dai_2021}, and in YbMgGaO$_4$ \cite{Shen2016, Shen2018}. This is a hallmark of a QSL with a spinon Fermi surface, as discussed in Sec. \ref{sec:QSL_signatures}. However, we also see that near $\boldsymbol{q}=\Gamma$ there is a small gap $\Delta\approx 0.25$ in the spectrum which is not seen in the experiments. If NaYbSe$_2$ is similar to KYbSe$_2$, which has been shown to be well modelled by a $J_1-J_2$ Heisenberg model with $J_1 \approx 0.56\,\mathrm{meV}$~\cite{Scheie2021Triangle}, then the gap would be $\Delta \approx 0.14\, \mathrm{meV}$, which is below the lowest frequency $\omega_{\mathrm{min}}\approx 0.2\, \mathrm{meV}$ accesible in Ref.~\cite{Dai_2021}. This discrepancy means that either these materials are not well modelled by the $J_1-J_2$ Heisenberg model, or that the lowest energies accessible in these experiments is not low enough to probe this gap on the order of $J_1 / 4$. In either case, the presence of the gap in the spectrum rules out the spinon Fermi surface state as ground state of this model.

Next we wish to distinguish the gapped $\mathbb{Z}_2$ from the gapless $U(1)$ Dirac spin liquid. To do this, we look at how the full spectrum changes as we tune $J_2$ through all three phases, illustrated in Fig. \ref{fig:J2_both}. As discussed in Sec. \ref{sec:QSL_signatures}, we want to look at what happens at $\boldsymbol{q}=K$ and $\boldsymbol{q}=M$, as we approach the QSL phase from the 120$^{\circ}$ phase. What we find is that there is a sharp low energy magnon branch near $\boldsymbol{q}=M$ that softens and decreases as the critical point is approached, and remains this way into the QSL phase. This feature has also been observed in a recent variational QMC study \cite{Ferrari2019}. This is a key signature of a gapless $U(1)$ Dirac QSL, suggesting the spectrum is gapless at both $\boldsymbol{q}=K$ and $\boldsymbol{q}=M$ in the QSL phase, in agreement with a recent DMRG study \cite{Hu_2019}. This feature is not captured within the Schwinger-boson formalism \cite{Scheie2021Triangle, Ghioldi_2022_SBT}, which finds that the gap $\boldsymbol{q}=M$ remains gapped as $J_2$ is tuned towards the critical point. This suggests that Schwinger-boson theory does not capture the QSL phase well, even though it has remarkable agreement with PEPS~\cite{Chi2022}, and the neutron spectrum of Ba$_3$CoSb$_2$O$_9$ \cite{Ito2017,Macdougal2020}, near the $J_2=0$ point.

To further examine the behavior at $\boldsymbol{q}=K$ and $\boldsymbol{q}=M$ across all three phases, we show $S(\boldsymbol{q}, \omega=0^+)$ as well in Fig. \ref{fig:J2_both}. To compute $S(\boldsymbol{q}, \omega=0^+)$ we integrate $S(\boldsymbol{q}, \omega)$ for $\omega_{\mathrm{min}} \le \omega \le \omega_{\mathrm{max}}$, with $\omega_{\mathrm{min}} = 0$, and $\omega_{\mathrm{max}} = 0.1$. We also adjusted the integration window, with $\omega_{\mathrm{min}} \in [0,0.1]$, and  $\omega_{\mathrm{max}} \in [0,0.2]$, as well as just using the value at $\omega=0$, with no qualitative difference. Thus, the low frequency cuts are robust and not just probing frequencies lower than can be resolved numerically. If this quantity is non-zero, then this would mean that there are gapless modes in the spectrum, which produce spectral weight down to the lowest energies. We find that deep in the 120$^{\circ}$ phase, that the low energy spectral weight is near zero, except at $\boldsymbol{q}=K$. As $J_2$ increases, low energy spectral weight develops at $\boldsymbol{q}=M$. Inside the QSL phase, we find appreciable weight at both $\boldsymbol{q}=K$ and $\boldsymbol{q}=M$, suggesting the spectrum is gapless at both these momenta, implying a gapless $U(1)$ Dirac spin liquid. Moreover, through all three phases, the spectral weight near $\boldsymbol{q}=\Gamma$ is zero, within machine precision. This provides further evidence against a spinon Fermi surface state, which would be non-zero everywhere within the QSL phase\cite{Savary2016}.

We want to note that separating zero from small is a challenging task numerically, and so we cannot definitively rule out the gapped $\mathbb{Z}_2$ or spinon Fermi surface spin liquid states. However, we believe the most likely interpretation of our data in the QSL phase is that the spectral weight is localized around $\boldsymbol{q}=K$ and $\boldsymbol{q}=M$, and broadened to yield non-zero weight along the full Brillouin zone boundary. These results are most consistent with a $U(1)$ Dirac QSL ground state.


\section{Conclusion}\label{sec:conclusions}

\subsection{Summary}

In this work we examine the dynamical structure factor for the full phase diagram of the $J_1-J_2$ Heisenberg model on the triangular lattice. We also look at the square lattice for the same model with $J_2=0$ as a benchmark for our method. For the square lattice, we compute the full spectrum, $S(\boldsymbol{q},\omega)$, and find great qualitative agreement with the results from QMC in Ref. \onlinecite{Shao2017}. From this, we extract the magnon dispersion relation $\epsilon(\boldsymbol{q})$ using Eq. \eqref{eq:w_max}, and then we compare this quantitatively with QMC~\cite{Shao2017} and SWT~\cite{PhysRevB.72.014403}. We find great agreement, except for wave-vectors $\boldsymbol{q}$ close to gapless modes. Nevertheless, we were able to extract the magnon velocity near $\boldsymbol{q}=M$ from the dispersion relation, and found excellent agreement with the velocity in Ref. \onlinecite{PhysRevB.92.195145}, which uses highly accurate methods based on static properties. The success on the square lattice provides confidence on utilizing this method to study the triangular lattice. 

In the $120^{\circ}$ magnetic long-range ordered phase, we compare our results against linear SWT~\cite{PhysRevB.79.144416}. We find low energy branches at frequencies much lower than the SWT prediction, at $\boldsymbol{q}=M$ and $\boldsymbol{q}=Y_1$. This reduction in the energy has been linked to avoided quasi-particle decay \cite{Verresen2019}, and also observed in variational QMC \cite{Ferrari2019}. This produces a roton-like mode at $\boldsymbol{q}=M$ which is not captured by linear SWT, and has been observed in neutron scattering experiments in Ba$_3$CoSb$_2$O$_9$~\cite{Ito2017,Macdougal2020}, and KYbSe$_2$~\cite{Scheie2021Triangle}, as well as simulations using PEPS~\cite{Chi2022}. Away from $\boldsymbol{q}=M$ and $\boldsymbol{q}=Y_1$, we also find good agreement with Schwinger-boson theory~\cite{Ghioldi_2022_SBT}. 

In the stripe ordered phase, we show the spectral function, dispersion relation, and velocity near $\boldsymbol{q}=M$ as a reference for future work. As far as we know, there are no other simulations of the full spectrum in this phase to compare against. In the static structure factor, we find ordering at $\boldsymbol{q}=M$, as observed previously \cite{PhysRevB.92.140403}. 

In the QSL phase, we find from the static structure factor that there is no unique ordering wave-vector. The weight is roughly evenly distributed on the path connecting $\boldsymbol{q}=K$ and $\boldsymbol{q}=M$, which are the ordering wave-vectors for the 120$^{\circ}$ and striped phase respectively. This demonstrates frustration in the system, as it struggles to satisfy simultaneously the $J_1$ and $J_2$ interactions, which is precisely what is expected to give rise to a QSL phase. 

We look for the signatures in the low energy spectrum predicted by the three dominant phases, which are summarized in Table \ref{tab:QSL_signatures}. We find a V-shape spectrum near $\boldsymbol{q}=\Gamma$, a key signature of a spinon Fermi surface. However, we find at low energies, a gap opens near $\boldsymbol{q}=\Gamma$, in contrast to the spinon Fermi-surface which is gapless everywhere \cite{Savary2016, PhysRevB.96.075105}. This V-shape pattern has been observed in neutron scattering experiments in NaYbSe$_2$~\cite{Dai_2021}, and in YbMgGaO$_4$~\cite{Shen2016, Shen2018}, which has been attributed to a spinon Fermi surface. If NaYbSe$_2$ is similar to KYbSe$_2$, which has been shown to be well modelled by a $J_1-J_2$ Heisenberg model with $J_1 \approx 0.56\,\mathrm{meV}$~\cite{Scheie2021Triangle}, then the gap would be $\Delta \approx 0.14\, \mathrm{meV}$, which is below the lowest frequency $\omega_{\mathrm{min}}\approx 0.2\, \mathrm{meV}$ accesible in Ref.~\cite{Dai_2021}. The absense of a low energy gap developing near $\boldsymbol{q}=\Gamma$ in these experiments means one of two things. Either, these materials are not well modelled by the $J_1-J_2$ Heisenberg model, and spin anisotropies or longer range interactions lead to different QSL phases than what is found in the $J_1-J_2$ Heisenberg model. Or, the gap appears at energies lower than what was accessible in these experiments. This calls for future neutron scattering experiments at lower energies, as well as ab-initio calculations, to elucidate the underlying microscopic models of these materials, and associated energy scales. 

As we increase $J_2/J_1$ up from $0$, we find that the low energy branch near $\boldsymbol{q}=M$ softens, and eventually leads to the gap closing at the critical point, ruling out a gapped $\mathbb{Z}_2$ QSL. This finding is consistent with the variational QMC study in Ref. \onlinecite{Ferrari2019}. We also find the high-energy branch near the $\Gamma$ point observed in their work. However, we find a continuum extending to higher energies, and this branch lies fully within this continuum as opposed to being isolated. We believe this difference is due to our method probing the full spectrum, and does not assume an ansatz for the excitations. The gap closing at $\boldsymbol{q}=M$ as the critical point is approached was not observed in the Schwinger-boson formalism \cite{Scheie2021Triangle, Ghioldi_2022_SBT}. We note that recent neutron scattering experiments in KYbSe$_2$ suggest it is well modelled by a $J_1-J_2$ Heisenberg model near the quantum critical point \cite{Scheie2021Triangle}. By applying hydrostatic pressure to KYbSe$_2$, and possibly the other triangular lattice materials, neutron scattering experiments may be able to give insights into what is happening near $\boldsymbol{q}=M$ as the material approaches quantum criticality.

Due to finite resolution in our simulations, we cannot definitively rule out a small gap at $\boldsymbol{q} = M$, which would imply a $\mathbb{Z}_2$ spin liquid, as recently claimed \cite{Scheie2021Triangle, Ghioldi_2022_SBT, Jiang_2022_arxiv}. We also cannot definitively distinguish zero spectral weight from small spectral weight near $\boldsymbol{q} = \Gamma$, as expected in a spinon Fermi surface state \cite{PhysRevB.96.075105, Savary2016}, and recently seen in neutron scattering experiments \cite{Dai_2021, Shen2016, Shen2018}. However, the most likely interpretation of our results is that there are gapless modes localized at $\boldsymbol{q}=K$ and $\boldsymbol{q}=M$, implying a gapless $U(1)$ Dirac spin liquid, in agreement with Ref. \onlinecite{Hu_2019}. 

\subsection{Perspectives}

Distinguishing the gapped $\mathbb{Z}_2$ from a gapless Dirac spin liquid remains a challenging task. Our results favor the gapless case, but future studies are needed to provide a definitive answer to this question. Perhaps looking at level crossings in the low energy spectrum could shed light on this, as was done for the $J_1-J_2$ Heisenberg model on the square lattice \cite{PhysRevLett.121.107202}, and the Shastry-Sutherland model \cite{PhysRevB.105.L060409,Wang_2022}.

We could also possibly gain insight to this question from cold atom experiments. Recently, a quantum spin liquid was realized in a quantum simulator of Rydberg atoms on the kagome lattice \cite{Semeghini_2021}. Triangular optical lattices have been proposed \cite{PhysRevB.100.140410},  and constructed \cite{PRXQuantum.2.020344,Yamamoto2020} to study frustrated quantum magnets. This may be a future direction to study the ground state properties of the $J_1-J_2$ Heisenberg model.

Quantum criticality can also be further explored by the Kibble-Zurek mechanism \cite{Kibble1976, Kibble1980, Zurek2014}. This procedure time evolves a ground state with a time-dependent Hamiltonian in proximity to a quantum critical point. Adiabaticity is lost as the gap closes, and the excitations generated are specified by the rate at which the Hamiltonian changes, and the critical exponents of the quantum critical point. Recent neutron scattering experiments in KYbSe$_2$ have found critical scaling with an unexplained critical exponent~\cite{Scheie2021Triangle}. Such a Kibble-Zurek process may be able to shed light on this observed criticality. Moreover, such Kibble-Zurek processes are ideal for use on quantum computers, on which unitary dynamics is easily programmed. If a quantum critical point is continuously connected to a product state, then a quantum computer may be useful in probing the critical point, as was done recently in the one-dimensional quantum Ising model \cite{PhysRevB.106.L041109}. Such small scale quantum computing devices are, in principle, not prone to the challenges of long-range interactions like MPS calculations are, for studying two-dimensional dynamics. Future work in this direction could provide insight into the nature of the QSL in the $J_1-J_2$ Heisenberg model, and simultaneously provide a problem where quantum computers extend beyond what is possible with classical simulations.

\section*{Note added}
While finalizing the current draft, we became aware of a similar work looking at the dynamical spin structure factor in the 120$^{\circ}$ and QSL phases of the $J_1-J_2$ Heisenberg model on the triangular lattice \cite{https://doi.org/10.48550/arxiv.2209.03344}.

\begin{acknowledgments}
    We gratefully acknowledge discussions with C.D. Batista, A. O. Scheie, D.A. Tennant, and M. Bintz on related works. We received support from the U.S. Department of Energy, Office of Science, Office of Basic Energy Sciences, Materials Sciences and Engineering Division under Award No. DE-AC02-05-CH11231 through the Theory Institute for Materials and Energy Spectroscopy (TIMES). J.E.M. was also supported by a Simons Investigatorship. This research used the Lawrencium computational cluster resource provided by the IT Division at the Lawrence Berkeley National Laboratory (supported by the Director, Office of Science, Office of Basic Energy Sciences, of the U.S. Department of Energy under Award No. DE-AC02-05CH11231). This research also used resources of the National Energy Research Scientific Computing Center (NERSC), a U.S. Department of Energy Office of Science User Facility operated under Award No. DE-AC02-05CH11231. 
\end{acknowledgments}

\bibliography{references}
\end{document}